\documentclass[prb,aps,amssymb,twocolumn,showpacs,floatfix]{revtex4}
\usepackage[dvips]{graphicx}
\renewcommand{\theequation}{\arabic{section}.\arabic{equation}}
\def\be{\begin{equation}}
\def\ee{\end{equation}}
\def\bea{\begin{eqnarray}}
\def\eea{\end{eqnarray}}
\def\ve{\varepsilon}

\def\bE{{\bf E}}

\def\X{{\cal X}}
\def\wc{\omega_c}
\def\ve{\varepsilon}

\def\w{\omega}
\def\St{{\rm St}}

\def\r{\mbox{\boldmath $r$}}
\def\q{\mbox{\boldmath $q$}}
\def\bzeta{\mbox{\boldmath $\zeta$}}

\begin{document}

\title{Microwave photoconductivity of a 2D electron gas:\\ 
Mechanisms and their interplay at high radiation power}
\author{I.A.~Dmitriev$^{1,*}$}
\author{A.D.~Mirlin$^{1,2,\dagger}$} 
\author{D.G.~Polyakov$^{1,*}$}
\affiliation{$^{1}$Institut f\"ur Nanotechnologie, Forschungszentrum
Karlsruhe, 76021 Karlsruhe, Germany\\
$^{2}$Institut f\"ur Theorie der kondensierten Materie,
Universit\"at Karlsruhe, 76128 Karlsruhe, Germany}

\date{\today}

\begin{abstract} We develop a systematic theory of microwave-induced
oscillations in the magnetoresistivity of a two-dimensional electron gas,
focusing on the regime of strongly overlapping Landau levels. At linear order
in microwave power, two novel mechanisms of the oscillations (``quadrupole''
and ``photovoltaic'') are identified, in addition to those studied before
(``displacement'' and ``inelastic''). The quadrupole and photovoltaic
mechanisms are shown to be the only ones that give rise to oscillations in the
nondiagonal part of the photoconductivity tensor. In the diagonal part, the
inelastic contribution dominates at moderate microwave power, while at
elevated power the other mechanisms become relevant. We demonstrate the
crucial role of feedback effects, which lead to a strong interplay of the four
mechanisms in the nonlinear photoresponse and yield, in particular, a
nonmonotonic power dependence of the photoconductivity, narrowing of the
magnetoresonances, and a nontrivial structure of the Hall photoresponse. At
ultrahigh power, all effects related to the Landau quantization decay due to a
combination of the feedback and multiphoton effects, restoring the classical
Drude conductivity. \end{abstract}


\maketitle

\section{Introduction}
\label{s1}
\setcounter{equation}{0}

In recent years, the nonequilibrium properties of quantum Hall systems in a
moderate perpendicular magnetic field $B$ have become a subject of intense
experimental\cite{zudov01,Ye01,yang02,mani02,zudov03,yang03,dorozhkin03,willett03,%
B-periodic,zudov04,du04,mani04,maniHall,Studenikin04,Kovalev,mani05,%
polarization,studenikin05,studenikin06,bykov1,bykov2,parr B,antidot,%
dorozhkinINTRA,bichrom,multi,dorozhkin06,inelasticDC,strongDC,ACDC} and
theoretical\cite{andreev03,ryzhii,durst03,dmitriev03,Raikh03,shi,ryzhii03,%
leiliu03,VA,short,classical,compress,michailov04,park,lee,Raikh04,Lyapilin1,%
ryzhii04,falkoprl,long,Halperin,Balents,vonoppen1,vonoppen2,torres05,%
leiliu05,Volkov,michailov04a,falkoprb,dietel,torres06,Kashuba06,Lyapilin2,%
leiliumulti,leiliubichrom,2componentDC,Auerbach,crossover} research (for
review see Refs.~\onlinecite{Fitzgerald,Girvin,Lyap,Durst06}). Much attention
has been attracted to the experimental discovery of a novel type of
$1/B$-periodic resistivity oscillations which arise in these systems under
microwave illumination.\cite{zudov01,Ye01} Remarkably, it was demonstrated
soon after the first experiments that at higher radiation power the minima of
the oscillations evolve into novel ``zero-resistance states''
(ZRS).\cite{mani02,zudov03}

As was shown in Ref.~\onlinecite{andreev03}, the ZRS can be understood as a
direct consequence of the oscillatory photoconductivity (OPC), provided the
latter may become negative. Independently of the microscopic origin of the
oscillations, the negative OPC signifies an instability leading to the
formation of spontaneous-current domains, which in turn yields the vanishing
of the observable resistance.

Unlike the explanation of the ZRS, so far there has been no common agreement
as to the microscopic origin of the OPC. The most frequently studied
mechanism\cite{ryzhii,durst03,shi,ryzhii03,leiliu03,VA,park,lee,Lyapilin1,%
ryzhii04,torres05,leiliu05,Volkov,torres06,Kashuba06,Lyapilin2,leiliumulti,%
leiliubichrom,Auerbach} of the OPC, suggested long ago in
Ref.~\onlinecite{ryzhii}, implies that the oscillations occur due to complex
scattering processes in which electrons simultaneously are scattered off
impurities, absorb (emit) microwave quanta, and are displaced along the
applied dc electric field. We term this mechanism a ``displacement''
mechanism. A different mechanism, called here the ``inelastic'' mechanism, was
recently proposed in Ref.~\onlinecite{dmitriev03} and studied in detail in
Refs.~\onlinecite{short,long} (similar ideas were also discussed in
Ref.~\onlinecite{dorozhkin03}). In the inelastic mechanism, the
magnetooscillations of the dc current are generated by a microwave-induced
change of the isotropic time-independent part $F_{00}(\ve)$ of the 
electron distribution function, 
\be
f(\ve,\varphi,t)=\sum_{\nu n} F_{\nu n}(\ve)\exp(i\nu\varphi+in\omega t)~,
\label{df}
\ee
 where $\ve$ is the electron energy, $\varphi$ the angle of the quasiclassical
momentum, $\omega$ the microwave frequency.

The above mechanisms have much in common. In both of them, the OPC originates
from oscillations of the density of states (DOS) $\nu(\ve)$ of
disorder-broadened Landau levels with changing $\ve$. Due to the oscillatory
$\nu(\ve)$, optical transitions in the system lead to a nonequilibrium
correction to the distribution function $f$ which oscillates with both
$\ve/\wc$ and $\omega/\wc$, where $\wc$ is the cyclotron frequency. In turn,
the $\omega/\wc$-oscillations of $f$ translate into the observed
$\omega/\wc$-oscillations of the magnetoresistivity.

A crucial difference between the two mechanisms is the following. In the
displacement mechanism, the radiation directly affects the first angular
harmonic $F_{10}(\ve)$ of the distribution function (and thus the dc current),
whereas all effects of the ac and dc fields on the even angular harmonics are
neglected. By contrast, in the inelastic mechanism the ac field influences the
isotropic part $F_{00}(\ve)$ only. A calculation of the dc response in the
resulting state with a modified $F_{00}(\ve)$ yields the OPC. While in the
first case the effect on the current is gained directly in every
microwave-assisted impurity-induced electron scattering event, in the latter
case the effect is accumulated over a long period of time during which the
electron diffuses in the field of impurities until it experiences inelastic
scattering. At experimentally relevant temperatures $T\sim 1$~K, the inelastic
relaxation is governed by electron-electron scatterings\cite{short,long} with
the inelastic relaxation time $\tau_{\rm in}\propto T^{-2}$.

In the linear regime with respect to the microwave power, the
contributions to the OPC generated by the displacement and inelastic
mechanisms sum up independently. A comparison shows [see Eqs.~(16) and
(17) of Ref.~\onlinecite{long} and Eqs.~(6) and (11) of
Ref.~\onlinecite{VA}] that both mechanisms produce the OPC with the
experimentally observed phase, period, and $B$-damping of the
oscillations. However, the slow inelastic relaxation rate (compared to
the rate of electron collisions with impurities, $\tau_{\rm q}^{-1}$)
makes the inelastic contribution larger by a factor $\tau_{\rm
  in}/\tau_{\rm q}\gg 1$. According to the calculation in
Ref.~\onlinecite{long}, $\tau_{\rm in}/\tau_{\rm q}\sim 100$ under the
experimental conditions.\cite{tauin} Apart from the magnitude of the
effect, the two contributions are qualitatively different in their
dependence on $T$ and polarization of the radiation. In accord with
the experiments, the inelastic contribution decreases as $\tau_{\rm
  in}\propto T^{-2}$ with increasing $T$ and does not depend on the
direction of linear polarization of the microwave field. By contrast,
the displacement mechanism yields a $T$ independent contribution which
depends essentially on the relative orientation of the microwave and
dc fields. Despite the above arguments, in a number of recent works
\cite{park,lee,Lyapilin1,ryzhii04,torres05,leiliu05,Volkov,torres06,%
  Kashuba06,Lyapilin2,leiliumulti,leiliubichrom,Auerbach} on this
subject the inelastic mechanism has not been taken into account.

In this work, we perform a systematic study of all relevant contributions to
the OPC both in the linear regime and in several nonlinear regimes which
emerge with increasing microwave power. On the experimental side, this work
has been motivated by a number of interesting nonlinear phenomena observed in
recent experiments.\cite{bichrom,multi,dorozhkin06,inelasticDC,strongDC,%
ACDC} From the theoretical point of view, a unified approach to the problem is
necessary since at high microwave power the interplay of the inelastic and
displacement mechanisms becomes essentially important. We show that the
different mechanisms of the OPC strongly affect each other due to feedback
effects already at moderate power levels which are easily accessible in the
experiment. In particular, while in the linear regime the displacement and
inelastic contributions to the OPC add up in phase, in the regime of a
``saturated inelastic contribution'' (SIC) the displacement contribution
changes its sign, which leads to a nonmonotonic behavior of the photoresponse
(still dominated by the inelastic contribution) with growing microwave
power. In the ultrahigh power regime, all contributions to the dc conductivity
that are related to the Landau quantization decrease with increasing power due
to a competition between the feedback and multiphoton effects. As a result,
the oscillations of the photoresponse vanish and the classical Drude
conductivity tensor is restored in the limit of high radiation power.

On top of the nonlinear interplay between the inelastic and
displacement mechanisms, we find two novel mechanisms leading to the
OPC, ``quadrupole'' and ``photovoltaic''. In the quadrupole mechanism,
the microwave radiation leads to excitation of the second angular
harmonic $F_{20}$ of the distribution function. The dc response in the
state with nonzero $F_{20}$ yields an oscillatory contribution to the
Hall part of the photoconductivity tensor which violates Onsager
symmetry. In the photovoltaic mechanism, a combined action of the
microwave and dc fields produces non-zero temporal harmonics $F_{21}$
and $F_{01}$, so that the distribution function (\ref{df}) acquires
oscillatory time dependence. The ac response in the resulting state
with excited $F_{21}$ and $F_{01}$ contributes to both the
longitudinal and Hall parts of the OPC. Provided $\tau_{\rm
in}/\tau_{\rm q}\gg 1$, the inelastic mechanism still gives the
dominant contribution to the diagonal part of the photoconductivity
tensor. However, the quadrupole and photovoltaic mechanisms are the
only ones yielding oscillatory corrections to the Hall part.

The paper is organized as follows. In Sec.~\ref{s2} we formulate the
problem and overview the main steps in the derivation of the quantum
kinetic equation obtained in Ref.~\onlinecite{VA}. Using this kinetic
equation, in Sec.~\ref{s3} we put forward a general classification of
contributions to the photoconductivity at first order in the microwave
power. We discuss here four distinctly different mechanisms of the
OPC. In Sec.~\ref{s4} we write down the kinetic equation for the case
of strongly overlapping Landau levels, which we focus on hereafter. In
Sec.~\ref{s5} we calculate the linear photoconductivity. In
Secs.~\ref{s6} and \ref{s7} we analyze several nonlinear regimes the
system passes through with increasing microwave power. Namely, in
Sec.~\ref{s6} we study the SIC regime at moderate power, driven by
feedback effects. In Sec.~\ref{s7} we turn to the interplay of the
feedback and multiphoton effects in the regime of ultrahigh microwave
power. In Sec.~\ref{s8} we consider the possibility of experimental
observation of the nonlinear effects in the magnetoresistivity
measurements. In Sec.~\ref{s9} the range of applicability of the
theory is discussed. The main results are summarized in
Sec.~\ref{s10}.

\section{2D electron gas in a classically strong magnetic
field under microwave irradiation} 
\label{s2}
\setcounter{equation}{0}
Our theory is based on the approach to the problem of kinetics of a 2D
electron gas (2DEG) in the presence of magnetic field and radiation that was
developed by Vavilov and Aleiner.\cite{VA} This approach was used for a
systematic study of the displacement mechanism of the OPC in
Ref.~\onlinecite{VA}. It was further developed to treat the inelastic
mechanism in Ref.~\onlinecite{long}. Based on controllable approximations, the
approach allows us to analytically describe the nonlinear behavior of the
system under intense microwave radiation.

\subsection{Model and parameters}\label{ss21}

Let us first specify the model. The relation between the main parameters of
our theory, which is also satisfied for the characteristic parameters of the
2DEG studied in the experiments on the OPC, is as follows:
\be 
\ve_F\gg T\,,\,\omega\,,\,\omega_c\,,\,\tau^{-1}_{\rm q} \gg\tau^{-1}_{\rm
tr}\,,\, \tau^{-1}_{\rm in}~. 
\label{parameters} 
\ee 
Here $\ve_F$ is the Fermi energy, $T$ the temperature, $\omega$ the microwave
frequency, $\omega_c$ the cyclotron frequency, $\tau_{\rm q}$ and $\tau_{\rm
tr}$ the quantum (single-particle) and transport disorder-induced scattering
times, respectively, and $\tau_{\rm in}$ the inelastic relaxation time. The
conditions (\ref{parameters}) imply:
\begin{itemize} 
\item{quasiclassical kinetics: the 2D electron states that contribute to
transport belong to high Landau levels in a narrow energy strip of width $T$
around the Fermi level, $\ve_F\gg T, \omega_c$;}
\item{predominantly small--angle scattering (smooth disorder), $\tau_{\rm
tr}\gg \tau_{\rm q}$;} 
\item{classically strong magnetic field, $\omega_c \tau_{\rm tr}\gg 1$;}
\item{pronounced oscillations in the DOS due to the Landau
quantization, $\omega_c\tau_{\rm q}\sim 1$.}  
\end{itemize}
In the high-mobility structures used in the experiment a smooth random
potential in the plane of the 2DEG was created by remote donors separated from
the 2DEG by a spacer of width $\xi/2\gg k_F^{-1}$, where
$k_F$ is the Fermi momentum. The Fourier transform of the 
correlation function of the random potential,
\be
\label{disorder}
W(q)=W(0)\,e^{-q\xi}~,
\ee 
falls off exponentially for momentum transfers $q\gg \xi^{-1}$, leading to
$\tau_{\rm q}/\tau_{\rm tr}=(k_F\xi)^{-2}\ll 1$, where the quantum and
transport scattering times in zero magnetic field are given by
\bea
\label{tauq}
&&\frac{1}{\tau_{\rm  q}}=\int\limits^{\infty}_{0}
{dq\over 2\pi}\,{W(q)\over 2v_F}~,\\
\label{tautr}
&&\frac{1}{\tau_{\rm  tr}}=\int\limits^{\infty}_{0}
{dq\over 2\pi}\,{W(q)\over 2 v_F}\left(\frac{q}{2 k_F}\right)^2~.
\eea
Effects of the smooth disorder on the spectral and transport
properties of 2D electrons occupying high Landau levels can be
properly described by the self-consistent Born approximation
(SCBA),\cite{Ando} provided that $\xi$ does not exceed the magnetic
length\cite{RaSha} (which is the case for the relevant magnetic fields
in most of the experiments).

\subsection{Quantum Boltzmann equation}
\label{ss22}

Using the SCBA under the conditions listed in Eq.~(\ref{parameters}), Vavilov
and Aleiner derived the quantum Boltzmann equation (QBE),\cite{VA} which is
the starting point of our calculation. The derivation of the QBE includes a
few key steps which we highlight below.

\begin{itemize}
\item{Moving coordinate frame.}

The initial problem of electron kinetics in the quantizing magnetic field and
the external (microwave + dc) electric field $\bE(t)$ in the presence of
static impurities is reduced to a problem of kinetics in the presence of
``dynamic'' impurities, whose potential is $t$ dependent, by changing to a
moving coordinate frame $\r\rightarrow \r-\bzeta(t)$, where $\bzeta(t)$ obeys
\bea
&&\partial_t \bzeta(t)=
\left(\frac{\partial_t -\w_c \hat{\varepsilon}}{\partial_t^2+\w_c^2}\right)
\frac{e\bE(t)}{m_e}~, \nonumber \\
&&\hat{\varepsilon}_{xy}=-\hat{\varepsilon}_{yx}=1~.
\label{zeta}
\eea 
All effects of the external electric field are now included in the time
dependence of $\bzeta(t)$. The transformation (\ref{zeta}), which is
particularly convenient for treating the external electric field
nonperturbatively, is equivalent to the transformation to Floquet states used
in a number of
works\cite{park,lee,Lyapilin1,torres05,Volkov,Kashuba06,Lyapilin2,Auerbach} on
the OPC.

\item{Keldysh equations within the SCBA.}

Within the SCBA, the Green's functions $\hat{G}^{(\alpha)}$ and the
self-energies $\hat{\Sigma}^{(\alpha)}$ [$(\alpha)\to R, A, K$ refers to the
retarded, advanced, and Keldysh components, respectively] are related to each
other as
\bea
\hat{\Sigma}^{(\alpha)}_{21}&=&\int\! \frac{d^2q}{(2\pi)^2}\,W(q)\,
e^{-i\q\bzeta_{21}}
\nonumber\\&\times&\left[e^{i\q\hat{\r}}\hat{G}^{(\alpha)} 
e^{-i\q\hat{\r}}\right]_{21}~,
\label{SCBAgen}
\eea
where the subscript (21) denotes the times $t_2$ and $t_1$ on the
Keldysh contour and $\bzeta_{21}=\bzeta(t_2) - \bzeta(t_1)$.

\item{Quasiclassical approximation.} 

\noindent Most importantly, using the conditions (\ref{parameters}),
Vavilov and Aleiner reduced Eq.~(\ref{SCBAgen}) to a simpler
quasiclassical equation in the ``action-angle'' representation:
\bea
\label{SCBA}
&&\Sigma^{(\alpha)}_{21}(\varphi)= 
-i\hat{\cal K}_{21}\,g^{(\alpha)}_{21}(\varphi)~,\\
\label{g}
&& g^{(\alpha)}_{21}(\varphi)\equiv i\wc\sum_k
\hat{G}^{(\alpha)}_{21}(\hat{n}+k;\,\hat{\varphi})~,
\eea
where the operators $\hat{n}$ and $\hat{\varphi}$ are canonically conjugated,
$[\hat{n},\hat{\varphi}]=-i$. The eigenvalues of $\hat{n}$ and $\hat{\varphi}$
are the Landau level index and the angle coordinate of the momentum,
respectively. The explicit expression for the linear operator $\hat{\cal K}$
is given below in Sec.~\ref{ss23}.

\end{itemize}
A significant difference between Eqs.~(\ref{SCBAgen}) and (\ref{SCBA}) is that
for high Landau levels the self-energy becomes $\hat{n}$ independent. It
follows that the distribution function $\hat{f}$, defined in the usual way by
\be\label{f}\hat{G}^R-\hat{G}^A-\hat{G}^K=2\,[\,\hat{G}^R
\hat{f}-\hat{f}\hat{G}^A\,]~,
\ee
commutes with $\hat{\varphi}$. Accordingly, the operator $\hat{\varphi}$,
which enters Eq.~(\ref{SCBA}) and the impurity collision integral
\be
\label{St}
i\,\St_{\rm im}\{ f \}=\hat{\Sigma}^R\hat{f}-\hat{f}\hat{\Sigma}^A+
{1\over 2}(\hat{\Sigma}^K+\hat{\Sigma}^A-\hat{\Sigma}^R)~, 
\ee
can be treated as a $c$--number. Substitution of Eqs.~(\ref{SCBA}) and
(\ref{g}) into Eqs.~(\ref{f}) and (\ref{St}) leads to the following quantum
kinetic equation:
\bea
\label{kineq}
{\cal L}\{f\}&=&\St_{\rm im}\{f\}\\\label{L}
{\cal L}\{f\}_{21}&=&(\partial_t+\wc\partial_\varphi)f_{21}-
\St_{\rm in}\{f\}_{21}~,\\
\St_{\rm im}\{f\}_{21}&=&\int\! dt_3\, \left[\,\hat{\cal K}_{21}(g_{23}^R
f_{31}-f_{23}g_{31}^A)\right.\nonumber \\ 
&-&\left. f_{31}\hat{\cal  K}_{23}g_{23}^R
+f_{23}\hat{\cal K}_{31}g_{31}^A\,\right]~.  
\label{St_im}
\eea
Here 
\be
t=(t_1+t_2)/2
\label{2.15}
\ee
is the ``center-of-mass'' time. The inelastic collision integral
$\St_{\rm in}\{f\}_{21}$ accounts for electron--electron scattering and for
coupling to a thermal (phonon) bath.

The spectrum of the problem is found from the coupled set of Eq.~(\ref{SCBA}) 
and the equation of motion
\bea
\nonumber
&&\left(i{\partial\over{\partial t_2}}-\wc\hat{n}\right)
\hat{G}^R_{21}(\hat{n},\,\hat{\varphi})=\frac{\delta(t_2-t_1)}{2\pi}
\\&&+\int_{t_1}^{t_2}\!dt_3\,\hat{\Sigma}^R_{23}(\hat{\varphi})
\,\hat{G}^R_{31}(\hat{n},\,\hat{\varphi})~.
\label{GR}
\eea
Altogether, Eqs.~(\ref{SCBA}), (\ref{kineq}), and (\ref{GR}) determine
both the spectrum and the distribution function. The dc current in the
presence of microwave radiation is then given according to Eq.~(3.30) of
Ref.~\onlinecite{VA} by 
\bea
\label{j} 
&&\overline{\bf j}=\overline{\bf j}_d+\overline{\bf j}_{nd}~,\\
\label{jnd}
&&\overline{\bf j}_{nd}= e n_e \,\overline{\partial_t \bzeta(t)}~,\\
&&\overline{\bf j}_d= -2 e p_F \left\langle{\cos\varphi\choose\sin\varphi}\,
2{\rm Re}\int\!d\tau \,
\overline{g^R_{t,\,t+\tau}f_{t+\tau,\,t}}\right\rangle~,
\nonumber\\
\label{curgen} 
\eea 
where ${\bf j}_{nd}$ and ${\bf j}_d$ stand for the nondissipative and
dissipative components of the current, respectively, and $n_e$ is the electron
concentration. Here and below, the bar denotes time averaging over the period
of the microwave field in the stationary state, the angular brackets denote
averaging over the angle $\varphi$.

\subsection{Kernel of the QBE}
\label{ss23}

To complete the formulation of the model, we now specify the kernel
${\hat{\cal K}}$ in Eqs.~(\ref{SCBA}) and (\ref{kineq}). According to
Ref.~\onlinecite{VA}, it is can be represented as
\bea
{\hat{\cal K}}_{21}&=&{1\over 2\tau_{\rm q}}
\int\limits_{-\infty}^\infty \!dx\,
\exp\left(\frac{x\partial_\varphi}{2k_F\xi}\right)\nonumber\\
&\times&e^{-|x|-i x \X_{21}(\varphi)}
\exp\left(\frac{x\partial_\varphi}{2k_F\xi}\right)~,
\label{kernel}
\eea
with all effects of the external electric field $\bE(t)$ being incorporated in
the function $\X_{21}(\varphi)\propto\bzeta_{21}\propto\bE$. Specifically, we
parametrize $\bE(t)$ as
\be
\label{E}
\bE(t)=\bE_{\rm dc}+\frac{E_\omega}{\sqrt{2}}\,{\rm Re}\left[\,{s_- 
+ s_+\choose i s_-- i s_+}e^{-i\omega t}\,\right]~,
\ee
where $\bE_{\rm dc}$ is the dc electric field, $E_\omega$ the
amplitude of the microwave field ${\bf E}_\omega$, and real $s_\pm$
with $s_+^2+s_-^2=1$ define elliptical polarization of ${\bf
  E}_\omega$ with the main axes along $x$ and $y$. In particular,
$s_+=1$ ($s_-=1$) corresponds to passive (active) circular
polarization, $s_+=s_-=1/\sqrt{2}$ to linear polarization along the
$x$--direction ($\varphi=0$).

Using Eqs.~(\ref{zeta}) and (\ref{E}), $\X\equiv\X_{21}(\varphi)$ is
written as 
\bea
\X&=&\frac{\bzeta(t_2)-\bzeta(t_1)}{\xi}{-\sin\varphi\choose\cos\varphi}
\nonumber\\ 
&=&\X_{\rm dc}+\X_\omega~, 
\label{X} 
\eea 
where 
\bea 
\label{Xdc}
&&\X_{\rm dc}=-t_-\,\sqrt{\frac{\tau_{\rm q}}{\tau_{\rm tr}}}\, \frac{e
v_F}{\wc } \,\bE_{\rm dc}{\cos\varphi\choose\sin\varphi}~,\\
\label{Xw}
&&\X_\omega=-\sin\frac{\omega t_-}{2}\sum_\pm\; {\cal
E}_\pm\cos(\varphi\pm\omega t)~, 
\eea 
$t_- =t_2-t_1$, $t$ is given by Eq.~(\ref{2.15}), and 
\be
{\cal E}_\pm=s_\pm\,\sqrt{\frac{2\tau_{\rm q}}{\tau_{\rm tr}}}\, \frac{e
E_\omega v_F}{\omega(\wc\pm\omega)}~.  
\label{Epm} 
\ee 
After a series expansion in $1/k_F \xi\ll 1$ the kernel (\ref{kernel})
acquires the final form we use below throughout the paper:
\bea 
{\cal K}&=&\tau_{\rm
q}^{-1}+{\cal K}_\bot +{\cal K}_j+{\cal K}_\varphi+{\cal O}[(k_F\xi)^{-2}]~,
\label{K}\\ 
\tau_{\rm q}{\cal K}_\bot&=&-{\X^2\over 1+\X^2}~, 
\label{Kbot}\\
\tau_{\rm q}{\cal K}_j&=& -i\sqrt{\frac{\tau_{\rm q}}{\tau_{\rm tr}}}
\left[\,\partial_\varphi\,\frac{\X}{(1+\X^2)^2}
+\frac{\X}{(1+\X^2)^2}\,\partial_\varphi\,\right]~,
\nonumber
\\ \label{Kj} 
\\ 
\tau_{\rm q}{\cal K}_\varphi&=& \frac{\tau_{\rm q}}{\tau_{\rm tr}}
\,\partial_\varphi\,\frac{1-3\X^2}{(1+\X^2)^3}\,\partial_\varphi~. 
\label{Kfi}
\eea
Equations (\ref{K})--(\ref{Kfi}) correspond to Eqs.~(3.48) and (3.49) of
Ref.~\onlinecite{VA}. We now turn to a systematic analysis of the QBE,
starting with a general classification of various contributions to the OPC at
order ${\cal O}(E_\omega^2)$.

\section{Mechanisms of the oscillatory photoconductivity}
\label{s3}
\setcounter{equation}{0}

In this section we present a general solution to the QBE (\ref{kineq}) to
first order in both the microwave power and the dc electric field. The case of
strongly overlapped Landau levels, $\wc\tau_{\rm q}\ll 1$, on which we focus
in this paper will be considered at this order in more detail in
Sec.~\ref{s5}. Throughout the paper we neglect any effect of the microwave
radiation on the functions $g^{R(A)}$ which determine the DOS:
\be
\label{nu}
\nu(\ve,t)=2\nu_0\,{\rm Re}\int\!\frac{d\varphi}{2\pi}\int\!dt_- 
\,e^{i\ve t_-} g^R_{t+t_-/2,\,t-t_-/2}(\varphi)~.
\ee
This approximation is well justified in the limit $\wc\tau_{\rm q}\ll 1$.
However, at $\wc\tau_{\rm q}\sim 1$ the microwave and dc fields may lead to a
pronounced modification of the DOS (see Secs.~V and VI of
Ref.~\onlinecite{VA}). A manifestation of the effect of the external electric
fields on the DOS in the photoresponse will be discussed
elsewhere.\cite{separated}

\subsection{Classical limit}
\label{ss31}

To better understand the role of different terms in the kernel (\ref{K}), it
is instructive to look first at the QBE in the classical (with respect to the
magnetic field) limit $\wc\tau_{\rm q}\to 0$. In this limit, the DOS
$\nu=\nu_0=m/2\pi$ is constant, the functions
$g^R_{21}=-g^A_{21}=\delta(t_2-t_1)/2$, and the collision integral
$\St_{\rm im}\{f\}_{21}
=\hat{\cal K}_{21}f_{21}-f_{21}K_0$, where $K_0=\tau_{\rm
q}^{-1}$ results from the action of the operator $\hat{\cal K}_{tt}$ on
unity. After the Wigner transformation, Eq.~(\ref{kineq}) reduces to
\bea
&&(\partial_t+\wc\partial_\varphi)f-\St_{\rm in}\{f\}\nonumber\\
&&=-\frac{f(\ve)}{\tau_{\rm q}}
+\int\frac{d\Omega}{2\pi}\,  
\hat{K}(\Omega)f(\ve -\Omega)~.
\label{Boltzmann}
\eea
In the absence of electric fields ($\X=0$), the r.h.s.\ of
Eq.~(\ref{Boltzmann}) is written as $K_\varphi f=\tau_{\rm
tr}^{-1}\partial_\varphi^2 f$, which describes a diffusive relaxation of the
electron momentum in a smooth random potential. This part of the kernel, which
is the smallest one in the parameter $1/k_F \xi\ll 1$, plays no role in this
paper.

The linear response to a small electric field $\X\ll 1$ is fully governed by
the part $\hat{{\cal K}}_j$ of the kernel, which is odd in
$\bE(t)$. Substituting ${\hat{\cal K}}_j$ into Eq.~(\ref{Boltzmann})
immediately gives the Drude formula for the conductivity. The term ${\cal
K}_\bot$, of zero order in $1/k_F \xi\ll 1$, is even in the electric field. It
modifies the isotropic part of the distribution function, which leads to the
effect of heating. Also, $\hat{{\cal K}}_\bot$ is responsible for the
electric-field-induced changes in higher even angular harmonics of $f$ (as we
show below, this yields additional contributions to the photoconductivity).

It is worth noting that under the standard assumption of an
energy--independent $\tau_{\rm tr}$ in the degenerate Fermi gas, all
nonlinear (in $\bE_{\rm dc}$ and $\bE_\omega$) corrections to the
Drude formula vanish to zero at the classical
level.\cite{classical,VA} In order to get a finite dc photoresponse,
one should then include an energy variation of $\tau_{\rm tr}$ around
the Fermi level, which is, however, still insufficient to get the
oscillatory photoresponse.\cite{classical} As far as the OPC 
is concerned, at the classical level it can only come
from the non--Markovian electron dynamics in a random potential
(``memory effects''). This classical contribution to the OPC (in
contrast to the classical contribution to the oscillatory ac
conductivity) was shown\cite{classical} to be parametrically smaller
than the quantum contributions we consider below.

\subsection{Quantum mechanisms  of the oscillatory photoconductivity}
\label{ss32}

The Landau quantization leads to a periodic modulation of the DOS for high
Landau levels, $\nu(\ve)=\nu(\ve +\wc)$, which, as far as the OPC is
concerned, essentially modifies the classical picture above. In particular, in
the high-temperature limit ($T\gg\omega, \wc$), the distribution function
acquires fast energy oscillations with the period $\omega_c$ on top of a
smooth thermal smearing (see Appendix~\ref{A1}):
\bea
f(\ve,t,\varphi)&=&f_T(\ve)+{\cal F}(\ve,t,\varphi)\,\partial_\ve f_T(\ve)~,
\nonumber\\
{\cal F}(\ve,t,\varphi)&=&{\cal F}(\ve+\wc,t,\varphi)~,
\label{fosc}
\eea 
where $f_T(\ve)$ describes the thermal distribution. What is important to us
is that the amplitude of the part of ${\cal F}$ that oscillates with $\ve$
oscillates also with the ratio $\omega/\wc$, which gives rise to the OPC.

Neglecting the effect of microwaves on the DOS, we thus assume that the
function $g^R_{t+\tau,\, t}(\varphi)$ does not depend on $t$ and
$\varphi$. Then, using Eq.~(\ref{nu}), the dissipative dc current
(\ref{curgen}) can be expressed as
\bea
\nonumber
\overline{{\bf j}_d}
&=& 2e v_F \left\langle{\cos\varphi\choose\sin\varphi}
\int\!d\ve\,\nu(\ve)\,\overline{f(\ve)}\right\rangle\\
&=& 2 e v_F \int\!d\ve\,\nu(\ve)\,{{\rm Re}\, F_{10}\choose-{\rm Im}\,
F_{10}}~,
\label{cur}
\eea
where $F_{10}(\ve)$ is the time-independent first angular harmonic of the
distribution function (\ref{df}). 

In the absence of external fields, 
\be
\label{Stph} 
\St^{(e-ph)}_{\rm in}\{f(\ve)\}=0
\ee
and the 2DEG is in equilibrium with a thermal bath at temperature $T$,
$f(\ve)=f_T(\ve)$. Applying the electric field produces perturbative
corrections to $f_T$:
\be
\label{expansionf}
f=f_T+\sum\limits_n\left({\cal L}^{-1}\St_{\rm im}\right)^n f_T~,
\ee
where ${\cal L}^{-1}=(\partial_t+\wc\partial_\varphi-\St_{\rm in})^{-1}$ is
the propagator in the l.h.s.\ of the QBE (\ref{kineq}) and the collisions
integral $\St_{\rm im}$, given by Eq.~(\ref{St_im}), should be expanded in
powers of $\bE_{\rm dc}$ and $\bE_\omega$ according to
Eqs.~(\ref{X})--(\ref{Kj}):
\bea
\St_{\rm im}&=&\sum\limits_{mk}\,\St^{(a,m,k)}_{\rm im}~,\nonumber\\
\St^{(a,m,k)}_{\rm im}&\propto&
\hat{\cal K}_a^{(m,k)}\propto\X_{\rm dc}^m\X_\omega^k~.  
\label{expansionSt}
\eea
The index $a=\bot,j$ (introduced for ease of visualization) shows which part
of the kernel, $\hat{\cal K}_j$ or $\hat{\cal K}_\bot$, enters $\St_{\rm im}$
at a given order in the fields: $a=\bot$ or $j$ for even and odd $m+k$,
respectively.

The condition $\tau_{\rm q}\ll\tau_{\rm tr}$ allows us to neglect $\hat{\cal
K}_\varphi$ and higher-order terms in the expansion of the kernel
(\ref{K}). To the same accuracy, we omit all terms in Eq.~(\ref{expansionf})
with $\hat{\cal K}_j$ entering more than once, since these are small in the
same parameter $\tau_{\rm q}/\tau_{\rm tr}$. The terms of zero order in
$\hat{\cal K}_j$ do not contribute to the current, as they are even in the
electric fields and, hence, in $\varphi$. As a result, we represent $F_{10}$,
which determines the dc current (\ref{cur}), in the form
\begin{widetext}
\be
F_{10}=\left\langle \overline{e^{-i\varphi}
\sum\limits_{\nu n}\sum\limits_{M+K={\rm odd}}
\left[\,{\cal L}^{-1}\!\sum\limits_{\mu+\chi={\rm even}}
\St_{\rm im}^{(\bot,\mu,\chi)}\,\right]^\nu 
{\cal L}^{-1}\,\St_{\rm im}^{(j,M,K)}
\left[\,{\cal L}^{-1}\!
\sum\limits_{m+k={\rm even}}\St_{\rm im}^{(\bot,m,k)}\,\right]^n 
f_T}\right\rangle~,
\label{expansionf10}
\ee
\end{widetext}
where the summation indices run over nonnegative values and $\mu+\chi$ and
$m+k$ are even, while $M+K$ is odd.

To obtain the linear dc response in the absence of radiation from
Eq.~(\ref{expansionf10}), we put $M=1$ and all other indices equal to zero,
which yields
\be
\label{fdc}
F_{10}=\left<\,\overline{ 
e^{-i\varphi}(\wc\partial_\varphi)^{-1}
\St_{\rm im}^{(j,1,0)}\{f_T\}}\,\right>~.
\ee
Here $\St_{\rm im}^{(j,1,0)}\{f_T\}$ is given by Eq.~(\ref{St_im}) with
$\hat{{\cal K}}$ represented by $\hat{{\cal K}}_j$, Eq.~(\ref{Kj}), and the
latter taken at first order in the dc field and at zero order in the microwave
field,
\be
\label{Kjdc}
\hat{\cal K}^{(1,0)}_j=-{i\over (\tau_{\rm q}\tau_{\rm
  tr})^{1/2}}\,(\partial_\varphi\X_{\rm dc}+\X_{\rm dc}
\partial_\varphi )~.
\ee
Being substituted into Eq.~(\ref{cur}), $F_{10}$ from Eq.~(\ref{fdc}) gives
the familiar linear-response dc conductivity, including the Shubnikov-de Haas
oscillations and the nonoscillatory quantum correction [see Eqs.~(\ref{Drude})
and (\ref{sigmaph1}) below].

\begin{figure}[ht]
\centerline{ 
\includegraphics[width=0.9\columnwidth]{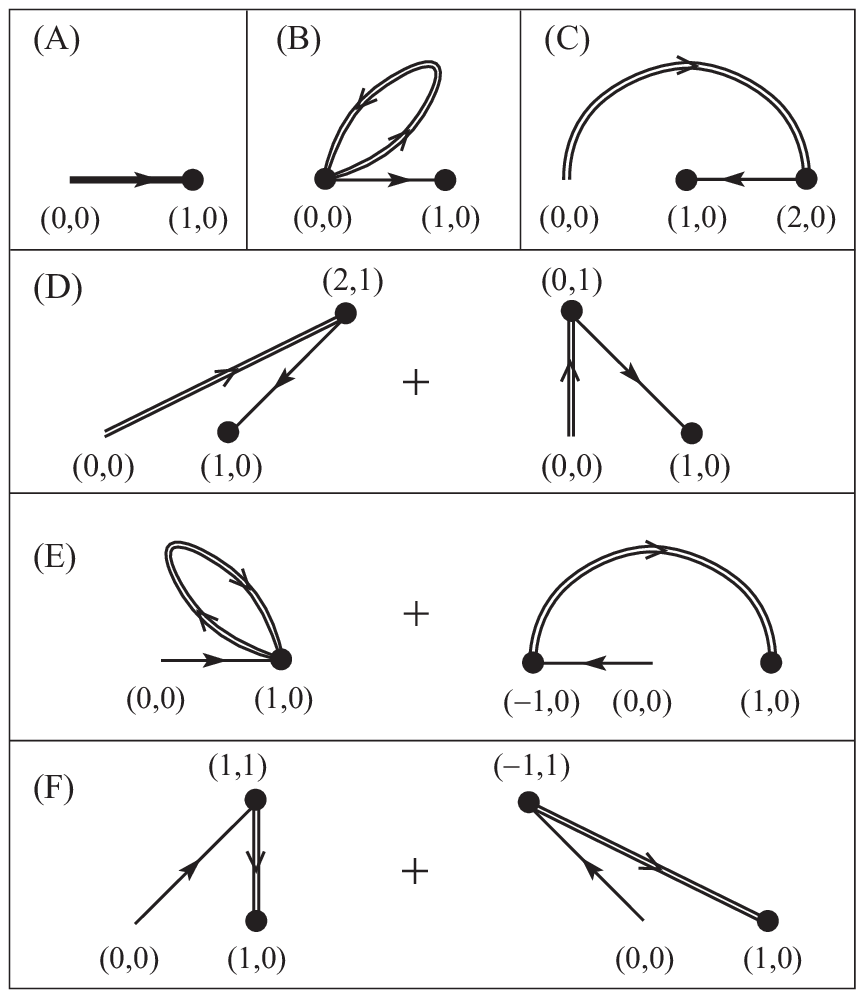}}
 
\caption{ All possible contributions to the photoconductivity at minimal order
   ${\cal O}(E_{dc} E_\omega^2)$ are illustrated as graphs in $(\nu,n)$ space
   of angular and temporal harmonics of the distribution function $F_{\nu n}$,
   according to Eqs.~(\ref{cur}) and (\ref{expansionf10}). The arrows denote
   coupling of the harmonics $F_{\nu n}$ by the impurity collision integral
   $\St_{\rm im}$ [Eq.~(\ref{St})]. The filled circles denote the action of
   the propagator ${\cal {\cal L}}^{-1}$ in the resulting state. More
   precisely, $\St_{\rm im}^{(j,1,2)}$ (which is of third order in the
   electric fields) is denoted by the thick line in diagram (A), while
   the first-order collision integrals $\St_{\rm im}^{(j,1,0)}$ and $\St_{\rm
   im}^{(j,0,1)}$ in diagrams (B)--(F) are denoted by the thin lines. The
   double-line arrows in diagrams (B)--(F) correspond to the second-order
   integrals $\St_{\rm im}^{(\bot,1,1)}$ and $\St_{\rm im}^{(\bot,0,2)}$. }
\label{diagrams}
\end{figure}

The effect of the microwave field on the dc current appears at order ${\cal
O}(E_{\rm dc} E_\omega^2)$. To this order, we pick up all terms with
$\mu\nu+M+mn=1$ and $\chi\nu+K+kn=2$ in the expansion (\ref{expansionf10}).
The result is illustrated in Fig.~\ref{diagrams}, where diagrams (A)--(F)
correspond to the following terms in Eq.~(\ref{expansionf10}):

\begin{itemize}
\item ``displacement'' contribution (A): 
$M=1$, $K=2$, $\nu=n=0$;
\item ``inelastic'' (B) and ``quadrupole'' (C) contributions: $M=n=1$,
  $k=2$, $K=\nu=m=0$. The two contributions are
  distinguished by the angular dependence of
  $\St_{\rm im}^{(\bot,0,2)}$: (B) and (C) include the isotropic and
  quadrupole parts of $\hat{{\cal K}}_\bot^{(0,2)}$, respectively;
\item ``photovoltaic'' contributions (D): 
  $K=n=m=k=1$, $M=\nu=0$;
\item  diagrams (E): $M=\nu=1$, $\chi=2$, $K=n=\mu=0$;
\item  diagrams (F):  $K=\nu=\mu=\chi=1$, $M=n=0$.
\end{itemize}
We now analyze these terms in more detail.

\subsubsection{Displacement mechanism (A)}

The displacement contribution to the distribution function is obtained by
expanding $\hat{{\cal K}}_j$ to order ${\cal O}(E_{\rm dc} E_\omega^2)$,
\be\label{KjA}
\hat{\cal K}^{(1,2)}_j=-{6 i\over (\tau_{\rm q}\tau_{\rm
  tr})^{1/2}}(\partial_\varphi\X_{\rm dc}
\X_\omega^2+\X_{\rm dc}\X_\omega^2\partial_\varphi)~,
\ee
and averaging $\X_\omega^2$ over $t$,
\be
\overline{\X_\omega^2}={1\over2}
({\cal E}_{+}^2+{\cal E}_{-}^2+2{\cal E}_{+}{\cal E}_{-}\cos2\varphi)
\,\sin^2\frac{\omega t_-}{2}~,
\label{X2}
\ee
while neglecting all effects related to $\hat{{\cal K}}_\bot$ by putting
$\nu=n=0$ in Eq.~(\ref{expansionf10}). The result,
\be\label{fA}
F^{(A)}_{10}=\left<\, e^{-i\varphi}(\wc\partial_\varphi)^{-1}
\,\overline{\St_{\rm im}^{(j,1,2)}\{f_T\}}\,\right>~,
\ee
is represented in Fig.~\ref{diagrams} by diagram (A). The thick arrow
denotes the action of the term in $\hat{\cal K}^{(1,2)}_j$ proportional to
$\exp (i\varphi)$, which couples the zero angular harmonic $f_{00}=f_T$ to
$f_{10}$. The filled circle stands for the propagator ${\cal
L}^{-1}=(\wc\partial_\varphi)^{-1}$ in the resulting state (1,0).


Equation (\ref{fA}) can be understood as follows. In the absence of the
microwave field, the dc current occurs due to a difference in the rates of
disorder-induced scattering with electron displacements along and against the
applied dc field, so that the distribution function becomes anisotropic,
Eq.~(\ref{fdc}). Absorption or emission of microwave quanta {\it during} these
scattering processes, Eq.~(\ref{fA}), modifies the current by changing the
final states and their occupancy. The displacement mechanism of the OPC was
predicted long ago in Ref.~\onlinecite{ryzhii} and addressed in the majority
of theoretical works\cite{ryzhii,durst03,VA,shi,ryzhii03,leiliu03,park,%
lee,Lyapilin1,ryzhii04,torres05,leiliu05,Volkov,torres06,Kashuba06,Lyapilin2,
leiliumulti,leiliubichrom} on the OPC. Its comprehensive study is presented in
Ref.~\onlinecite{VA}.

\subsubsection{Inelastic mechanism (B)}

The inelastic contribution to the photoconductivity is governed by a change of
the isotropic time--independent part of the distribution function, $F_{00}$, due to
the absorption and emission of microwave quanta. The change of $F_{00}$ is
related to the $t$ and $\varphi$ independent part of the kernel, which, at
order ${\cal O}(E_\omega^2)$, is given by
\bea
\langle\overline{{\cal  K}^{(0,2)}_\bot}\rangle&=&
-{1\over\tau_{\rm q}} \left<\overline{\X_\omega^2}\right>\nonumber\\
&=&-\frac{1}{2\tau_{\rm q}}({\cal E}_{+}^2+
{\cal E}_{-}^2)\,\sin^2\frac{\omega t_-}{2}~.
\label{KbotB}
\eea
The QBE for $F_{00}$ to order ${\cal O}(E_\omega^2)$ reads
\bea
\St^{(e-ph)}_{\rm in}\{F_{00}\}+\St^{(e-e)}_{\rm in}\{F_{00}\}
=-\left<\,\overline{\St_{\rm im}^{(\bot,0,2)}\{f_T\}}\,\right>~.\nonumber\\
\label{kineq00}
\eea
Equation (\ref{kineq00}) with Eq.~(\ref{KbotB}) substituted into the r.h.s.\
describes two effects: (i) electron heating, i.e., a modification of the
smooth part of the electron distribution, and (ii) the appearance of an
oscillatory correction to the isotropic part of the distribution on
top of the smooth part [cf.\ Eq.~(\ref{fosc})]. The heating is controlled by
electron--phonon scattering [the first term on the l.h.s. of
Eq.~(\ref{kineq00})] and leads to an effective electronic temperature $T_e$
larger than the bath temperature $T$. We relegate the discussion of $T_e$ as a
function of the system parameters and the microwave power to Sec.~\ref{s8}.
Much faster processes of electron-electron inelastic scattering (the second
term on the l.h.s.) cannot stabilize the increase of $T_e$, but are capable of
equilibrating electrons among themselves. The amplitude of the part of $\cal
F$ oscillating with $\ve$ in Eq.~(\ref{fosc}) and the amplitude of the
oscillations in the photoconductivity are thus proportional to the rate of
e--e collisions,\cite{long} $\tau_{\rm in}^{-1} =\tau_{\rm in}^{-1}(T_e)$,
\be
\label{tauin}
\St^{(e-e)}_{\rm in}\{F_{00}\}=-\tau_{\rm in}^{-1}F_{00}~.
\ee
Having obtained the nonequilibrium $F_{00}$, we use $\hat{{\cal K}}_j$ given by
Eq.~(\ref{Kjdc}) to calculate the linear response with respect to the dc
field, so that altogether at order ${\cal O}(E_{\rm dc}E_\omega^2)$ we have:
\be
F^{(B)}_{10}=\left<\, e^{-i\varphi}(\wc\partial_\varphi)^{-1}
\St_{\rm im}^{(j,1,0)}\left[
\tau_{\rm in}\left<\,\overline{\St_{\rm im}^{(\bot,0,2)}\{f_T\}}\,
\right>\,\right]\,\right>~.
\label{fB}
\ee
Equation (\ref{fB}) is represented by diagram (B) in Fig.~\ref{diagrams},
where the double--line loop denotes the action of $\langle\,\overline{\St_{\rm
im}^{(\bot,0,2)}}\,\rangle$ on $f_T$ and the thin arrow corresponds to the
action of $\St_{\rm im}^{(j,1,0)}$ on the resulting $F_{00}$.

One can see that, as compared to the displacement contribution to the OPC, the
effect of the inelastic mechanism is accumulated during a much longer time
$\tau_{\rm in} \gg\tau_{\rm q}$. The period and the phase of the oscillations
are the same for the two mechanisms (and correspond to those observed in the
experiment); however, the amplitude of the inelastic contribution is
$\tau_{\rm in}/\tau_{\rm q}\gg 1$ times larger. Another important difference
is that the $T$ dependence of $\tau_{\rm in}$ makes it possible to explain the
temperature-induced decay of the oscillations as observed in the experiment,
in contrast to the oscillations produced by the displacement mechanism, which
are $T$ independent. The inelastic mechanism of the OPC was proposed in
Ref.~\onlinecite{dmitriev03} and discussed in detail in
Ref.~\onlinecite{long}.

While mechanisms (A) and (B) are widely discussed in the literature,
already at first order in the microwave power there exist additional
contributions (C)--(F) to the OPC, which have not been studied before.

\subsubsection{Quadrupole mechanism (C)}

In addition to the effect on $F_{00}$, the absorption and emission of
microwave quanta, described by $\hat{{\cal K}}^{(0,2)}_\bot=-\tau_{\rm
q}^{-1}\X_\omega^2$, leads to the appearance of a quadrupole correction
$F_{20}$ to the distribution function [see Eq.~(\ref{X2})]:
\be
\label{f20}
F_{20}=\left< e^{-2 i\varphi}(\wc\partial_\varphi)^{-1}
\,\overline{\St_{\rm im}^{(\bot,0,2)}\{f_T\}}\right>~.
\ee
Acting by the operator $\hat{{\cal K}}^{(1,0)}_j$ [Eq.~(\ref{Kjdc})] on the
quadrupole correction yields
\be
\label{fC}
F^{(C)}_{10}=\left<\, e^{-i\varphi}(\wc\partial_\varphi)^{-1}
\,\St_{\rm im}^{(j,1,0)}\{e^{2i\varphi}F_{20}\}\,\right>~,
\ee
as illustrated by diagram (C) in Fig.~\ref{diagrams}.

\subsubsection{Photovoltaic mechanism (D)}

The photovoltaic contribution to the photoconductivity is generated by the
action of
\be
\label{KbotD}
\hat{{\cal  K}}^{(1,1)}_\bot=
-2 \tau_{\rm q}^{-1} \X_\omega \X_{dc}
\ee
on the equilibrium distribution, which leads to the excitation of the
components $F_{01}$ and $F_{21}$ oscillating with time. Acting then by
\be
\label{KjD}
\hat{{\cal  K}}^{(0,1)}_j=-i(\tau_{\rm q}\tau_{\rm
  tr})^{-1/2}(\partial_\varphi\X_\omega+\X_\omega\partial_\varphi)~,
\ee
we calculate the ac response in the resulting state, which gives
\bea
\nonumber
F^{(D)}_{10}&=&\left<\, e^{-i\varphi}(\wc\partial_\varphi)^{-1}\right.\\
&\times&\left.\overline{\St_{\rm im}^{(j,0,1)}
\left[\,
(\wc\partial_\varphi+\partial_t)^{-1}\,
\St_{\rm im}^{(\bot,\, 1,1)}\{f_T\}\,\right]}\,
\right>~,
\nonumber\\
\label{fD}
\eea
see diagrams (D) in Fig.~\ref{diagrams}

\subsubsection{``Inverse-order'' contributions (E) and (F)}

Additionally, contributions similar to (B)--(D) but with the inverse
order of the operators $\St_{\rm im}^{(j)}$ and $\St_{\rm
im}^{(\bot)}$ in Eq.~(\ref{expansionf10}) are also possible. Diagrams
(E) and (F) in Fig.~\ref{diagrams}, which describe these processes,
involve harmonics of the distribution function different from those in
diagrams (B)--(D).  Namely, while contributions (B)--(D) include even
angular harmonics only, except for the final $F_{10}$, all harmonics
in diagrams (E) and (F) are odd in the angle of the momentum, except
for the initial $F_{00}=f_T$.  Despite yielding nonzero contributions
to $F_{10}$, diagrams (E) and (F) vanish in the dc current in the
limit of high $T$ and strongly overlapping Landau levels [see the
paragraph following Eq.~(\ref{kinbot})]. The explicit calculation in
Secs.~\ref{s4} and \ref{s5} shows that, in the limit of strongly
overlapping Landau levels, mechanisms (A)--(D) give a complete set of
contributions to the photoresponse to leading order in the microwave
power.

\section{QBE for overlapping Landau levels}
\label{s4}
\setcounter{equation}{0}

From now on, we study the effect of microwave radiation on dc transport
in the case when the Landau quantization modulates the DOS only weakly. In
this limit, the modulation is represented by a single cosine term,
\be
\label{nuOvLL}
\tilde{\nu}(\ve)=\nu(\ve)/\nu_0=1-2\delta\,\cos\ve t_B~,\; t_B=2\pi/|\wc|~,
\ee
where 
\be
\delta=\exp(- t_B/2\tau_{\rm q})\ll 1~.
\label{delta}
\ee
It is worth mentioning that the modulation of the DOS at order ${\cal
O}(\delta)$ is insensitive to the external electric fields
[corrections\cite{VA,separated} to the DOS induced by the electric fields
appear at order ${\cal O}(\delta^2)$].

The existence of the small parameter $\delta$ greatly simplifies the solution
of the kinetic equation in energy space, see Appendix~\ref{A2}. The solution
can be represented in the form (\ref{fosc}) with
\bea
{\cal F}(\ve,t,\varphi)&=&\phi_0(\varphi,\,t)\nonumber\\
&-&2\delta\;{\rm Re}\,\left[\,\phi_1(\varphi,\,t)
\exp(i\ve t_B)\,\right]~.
\label{prephi}
\eea
For $T\gg t_B^{-1}$, the integration over $\ve$ in Eq.~(\ref{cur}) averages
out terms in the current of first order in $\delta$. As a result, in the
high-$T$ limit the oscillations of the DOS manifest themselves in the current
at order ${\cal O}(\delta^2)$. We therefore expand $\phi_0$ up to a term
quadratic in $\delta$:
\be
\phi_0=\phi_0^{\rm D}+2\delta^2\phi_0^{(2)}~,
\label{defphi0}
\ee
where $\phi_0^{\rm D}$ is the Drude part independent of $\delta$, and write
the current to order ${\cal O}(\delta^2)$ as
\be
\label{preOvcur}
\overline{{\bf j}_d}= 
2 e v_F \nu_0 {{\rm Re}\choose -{\rm Im}}
\left<\, e^{-i\varphi}\,
(\,\overline{\phi_0}+2\delta^2\,{\rm Re}
\,\overline{\phi_1}\,)\,\right>~.
\ee
It is convenient to introduce new functions $\phi_j$ and $\phi_\bot$ according
to
\bea
\label{defphij}
&&\phi_j=\phi_0^{(2)}+{\rm Re}\,\phi_1\,,\\
\label{defphibot}
&&\phi_\bot={\rm Im}\,\phi_1~,
\eea
after which the current (\ref{preOvcur}) is split into the classical
(Drude) and quantum parts as follows:
\be
\label{Ovcur}
\overline{{\bf j}_d}= 2 \sigma_{xx}^{\rm D}\bE_{\rm dc}+ 4 \delta^2
e v_F \nu_0 {{\rm Re}\choose {\rm Im}}
\left\langle\, e^{i\varphi}\,\overline{\phi_j}\,\right\rangle~.
\ee 
As shown in Appendix~\ref{A2}, the functions $\phi_0^{\rm D}$, $\phi_j$, and
$\phi_\bot$ obey
\bea
\label{kinD}
&&i(\partial_t+\wc\partial_\varphi)\phi_0^{\rm D}=
\partial_{t_-}\left.\hat{\cal K}_j(t,t_-)\right|_{t_-=0}~,\\
\nonumber
&&\left. i(\partial_t+\wc\partial_\varphi)\phi_j=
(\partial_{t_-}-{1\over2}\partial_t)
\hat{\cal K}_j(t,t_-)\right|_{t_-=t_B}\\
\label{kinj}
&&\qquad -\hat{\cal K}_j(t,t_B)\phi_\bot-\phi_\bot\hat{\cal K}_j(t,t_B)~,
\\
\nonumber
&&[\partial_t+\wc\partial_\varphi-\hat{\cal K}_\bot(t, t_B)]\phi_\bot
+\tau_{\rm in}^{-1}\left\langle\phi_\bot\right\rangle\\
\label{kinbot}
&&\left. \qquad =({1\over2}\partial_t-\partial_{t_-})
\hat{\cal K}_\bot(t,t_-)\right|_{t_-=t_B}~.
\eea
Here and below, any operator $\hat{{\cal K}}_{j,\bot}$ in the rightmost
position is understood as acting on unity.

The first term on the r.h.s.\ of Eq.~(\ref{kinj}) gives the displacement
contribution (A) to the dc current (\ref{Ovcur}), while the second and third
terms yield the inelastic, quadrupole, and photovoltaic contributions
(B)--(D).  It is important to notice that Eq.~(\ref{kinj}) for the function
$\phi_j$ includes the odd part of the kernel $\hat{\cal K}_j$ only, while the
function $\phi_\bot$ [Eq.~(\ref{kinbot})] is completely determined by the even
part $\hat{\cal K}_\bot$. It follows that diagrams (E) and (F) in
Fig.~\ref{diagrams} produce no contribution to the dc current, because of the
absence of coupling between $\phi_j$ and $\hat{\cal K}_\bot$. Although the
functions $\phi_0^{(2)}$ and ${\rm Re}\phi_1$ are coupled with each other [see
Eqs.~(\ref{F02}) and (\ref{ReF1})] and do have contributions of type (E) and
(F), each of these contributions vanishes in the combination $\phi_j$
[Eq.~(\ref{defphij})] which determines the current according to
Eq.~(\ref{Ovcur}).

As follows from Eq.~(\ref{Ovcur}), for the calculation of $\overline{{\bf
j}_d}$ we need only the $t$ independent parts of $\phi_0^{\rm D}$ and
$\phi_j$. Using the explicit form of $\hat{{\cal K}}_j$ [Eq.~(\ref{Kj})] and
the fact that $\X(t_-=0)=0$, the r.h.s.\ of Eq.~(\ref{kinD}) averaged over $t$
reads
\be
\label{preDrude}
\partial_{t_-}\!\left.\overline{\hat{\cal K}_j(t,t_-)}\right|_{t_-=0}
=-i\partial_\varphi\frac{\partial_{t_-}
\X_{\rm dc}}{\sqrt{\tau_{\rm q}\tau_{\rm tr}}}~,
\ee
which gives
\be
\label{phiDrude}
\overline{\phi_0^{\rm D}}=
\frac{e v_F}{\wc^2\tau_{\rm tr}} \bE_{\rm dc} 
{\cos\varphi\choose\sin\varphi}
\ee
and, correspondingly, the Drude conductivity
\be
\label{Drude}
\sigma_{xx}^{\rm D}=\frac{e^2\nu_0 v_F^2}{2\wc^2\tau_{\rm tr}}~.
\ee
Similarly,
\be
\label{phij}
\overline{\phi_j}=
\overline{\frac{-\partial_{t_B}+2\phi_\bot}
{\wc\sqrt{\tau_{\rm q}\tau_{\rm tr}}}\,\frac{X}{(1+X^2)^2}}~,
\ee
where
\be
X\equiv\X\vert_{t_-= t_B}=X_{\rm dc}+X_\omega
\ee
and for later use we introduce also
\bea
X_\omega&\equiv&\left.\X_\omega\right|_{t_-= t_B}~,\nonumber\\
X_{\rm dc}&\equiv&\left.\X_{\rm dc}\right|_{t_-= t_B}~.
\label{defX}
\eea
The derivative $\partial_{t_B}$ in Eq.~(\ref{phij}) should be understood
according to the notation which we also use below in the paper:
\be
\label{dtb}
\partial_{t_B}A(X_\omega,X_{\rm dc})
\equiv\partial_{t_-}A(\X_\omega,\X_{\rm dc})
\vert_{t_-= t_B}~,
\ee
where $A$ is an arbitrary function.

Now that $\overline{\phi_j}$, which determines the dc current (\ref{Ovcur}),
is obtained in the form of Eq.~(\ref{phij}), the problem of finding the OPC to
arbitrary order in $E_\omega^2$ or $E_{\rm dc}$ is reduced to solving
Eq.~(\ref{kinbot}) for the function $\phi_\bot$. In this paper, we restrict
ourselves to the calculation of the linear response with respect to the dc
field. In Sec.~\ref{s5}, we obtain explicit expressions for the linear
photoconductivity, corresponding to $\overline{\bf j}\propto E_{\rm dc}
E_\omega^2$. With increasing microwave power, the system passes through
several nonlinear regimes, which are studied in Secs.~\ref{s6}, \ref{s7}, and
\ref{s8}.

\section{Linear photoconductivity}
\label{s5}
\setcounter{equation}{0}

According to the general classification presented in Sec.~\ref{s3}, at order
${\cal O}(E_{\rm dc} E_\omega^2)$ one can distinguish four different
mechanisms of the photoconductivity, illustrated by diagrams (A)--(D) in
Fig.~\ref{diagrams}. We now calculate terms in the photoconductivity
associated with the corresponding nonequilibrium corrections to the
distribution function, given by Eqs.~(\ref{expansionf10}), (\ref{fA}),
(\ref{fB}), (\ref{fC}), and (\ref{fD}).

The displacement contribution (A), Eq.~(\ref{fA}), is produced by ${\cal
K}^{(1,2)}_j\propto E_{\rm dc} E_\omega^2$. All effects on even angular
harmonics of the distribution function, governed by $\hat{{\cal K}}_\bot$, are
neglected, i.e., $\phi_\bot^{(A)}=0$. Equation~(\ref{phij}) then reduces to
\be
\label{phiA}
\overline{\phi_j^{(A)}}=
\frac{6\,\partial_{t_B}}{\wc\sqrt{\tau_{\rm q}\tau_{\rm tr}}}
\,X_{\rm dc}\overline{X_\omega^2}~.
\ee

By contrast, contributions (B)--(D) are related to the effect of the microwave
and dc fields on even angular harmonics. At order ${\cal O}(X^2)$,
Eq.~(\ref{kinbot}) reads
\be
\label{kinbotX2}
(\partial_t+\wc\partial_\varphi)\phi_\bot
+\tau_{\rm in}^{-1}\left\langle\phi_\bot\right\rangle=
\tau_{\rm q}^{-1}(\partial_{t_B}-\partial_t/2)X^2~.
\ee
To linear order in the dc field, we replace $X^2$ in the r.h.s. of
Eq.~(\ref{kinbotX2}) by $X_\omega^2+2 X_\omega X_{\rm dc}$.  The isotropic and
quadrupole parts of $\overline{X_\omega^2}$ [Eq.~(\ref{X2})] yield the
inelastic (B) and quadrupole (C) contributions, respectively, while the
photovoltaic contribution (D) is generated by the product $X_\omega X_{\rm
dc}$:
\bea
\label{phiB1}
&&\phi_\bot^{(B)}=\frac{\tau_{\rm in}}{\tau_{\rm q}}
\partial_{t_B}\langle \overline{X_\omega^2}\rangle~,\\\label{phiC1}
&&\phi_\bot^{(C)}=\frac{\partial_{t_B}}{\wc\tau_{\rm q}\partial_\varphi}
\left(
\overline{X_\omega^2}-\langle \overline{X_\omega^2}\rangle
\right)~,\\
&&\phi_\bot^{(D)}=\tau_{\rm q}^{-1}(\partial_t
+\wc\partial_\varphi)^{-1}(2\partial_{t_B}-\partial_t)
X_\omega X_{\rm dc}~.\nonumber\\
\label{phiD1}
\eea
In Eqs.~(\ref{phiB1})--(\ref{phiD1}), $\phi_\bot^{(B)}$ and $\phi_\bot^{(C)}$
are averaged over $t$, whereas $\phi_\bot^{(D)}$ is not. The current
(\ref{Ovcur}) is then produced by the dc [in the case of (B) and (C)] or ac
[in the case of (D)] response in the resulting state:
\be\label{phijB-D}
\overline{\phi_j}=
\frac{2}{\wc\sqrt{\tau_{\rm q}\tau_{\rm tr}}}\,
\overline{\phi_\bot(X_{\rm dc}+X_\omega) }~.
\ee
Substituting Eqs.~(\ref{phiB1})--(\ref{phiD1}) into Eq.~(\ref{phijB-D}) and
using the resulting $\overline{\phi_j}$ together with
$\overline{\phi_j^{(A)}}$ [Eq.~(\ref{phiA})] in Eq.~(\ref{Ovcur}) yields the
current ${\bf \overline{j}}=\hat\sigma_{\rm ph}\bE_{\rm dc}$.

It is convenient to parametrize the photoconductivity tensor $\hat\sigma_{\rm
ph}$ by four functions $d_s$, $d_a$, $h_s$, $h_a$:
\bea
\nonumber
\frac{\hat{\sigma}_{\rm ph}}{2\sigma^{\rm D}_{xx}}&=&
\left(\!\begin{array}{cc}1+2\delta^2&-\wc\tau_{\rm tr}\\ 
\wc\tau_{\rm tr}&1+2\delta^2\end{array}\!\right)\\
&-&2\delta^2\left(\!
\begin{array}{cc}d_s & h_a \\ -h_a & d_s \end{array}\!\right)
-2\delta^2\left(\!
\begin{array}{cc}d_a & h_s \\ h_s & -d_a \end{array}\!\right)~.\nonumber\\
\label{sigmaph1}
\eea
The first matrix in Eq.~(\ref{sigmaph1}) is the linear dc conductivity in the
absence of microwaves. The quantum correction $2\delta^2$ to the dissipative
part comes from inserting Eq.~(\ref{phij}) into Eq.~(\ref{Ovcur}) if one
puts $\phi_\bot=0$ in Eq.~(\ref{phij}) and also substitute $X_{\rm dc}$ for
$X/(1+X^2)^2$. The resulting $\overline{\phi_j}$ coincides with $\phi_0^{\rm
D}$ [Eq.~(\ref{phiDrude})].

Two other matrices in Eq.~(\ref{sigmaph1}) represent the microwave--induced
corrections. We express the elements of these matrices in terms of the
following dimensionless parameters:
\bea
\label{w}
w&=&\omega t_B/2=\pi\omega/|\wc|~,\\\label{Q}
Q&=&\sin^2 w \,({\cal E}_{+}^2+{\cal E}_{-}^2)~,\\\label{QAI} 
Q_{\rm AI}&=&\sin^2 w \,{\cal E}_{+}{\cal E}_{-}~,\\
Q_{\rm S}&=&2\sin^2 w\nonumber\\
\label{QS}
&\times&\left({\cal E}_+^2\,\frac{\omega+\wc}{\omega+ 2\wc}
+{\cal E}_-^2\,\frac{\omega-\wc}{\omega- 2\wc}\right)~,\\
Q_{\rm AS}&=&2\sin^2 w \,{\wc^2\over\omega}
\nonumber\\
\label{QAS}
&\times&\left({\cal E}_+^2\,\frac{1}{\omega+ 2\wc}
+{\cal E}_-^2\,\frac{1}{\omega- 2\wc}\right)~,
\eea
where ${\cal E}_\pm$ are defined in Eq.~(\ref{Epm}).  In addition to
the effective microwave power $Q$ we introduced three similar
parameters, Eqs.~(\ref{QAI})-(\ref{QAS}), with the subscripts
``AI'',``S'', and ``AS'' standing for ``anisotropic'', ``symmetric'',
and ``antisymmetric'', respectively (in accord with their appearance
in the corresponding components of $\hat\sigma_{\rm ph}$, see below).

The diagonal isotropic part $d_s$ of $\hat\sigma_{\rm ph}$,
Eq.~(\ref{sigmaph1}), is a sum of the displacement, inelastic, and
photovoltaic contributions: 
\bea\label{ds1}
d_s&=&d_s^{(A)}+d_s^{(B)}+d_s^{(D)}~,\\\label{dsA1}
d_s^{(A)}&=&3Q+6Q\,w\cot w~,\\\label{dsB1} 
d_s^{(B)}&=&2{\tau_{\rm in}\over\tau_{\rm q}}Q\,w\cot w~,\\\label{dsD1} 
d_s^{(D)}&=&-{w\,Q_{\rm S}\over\omega\tau_{\rm q}}~.
\eea
The diagonal anisotropic part $d_a$ is governed by the displacement and
photovoltaic mechanisms:
\bea
\label{da1}
d_a&=&d_a^{(A)}+d_a^{(D)}~,\\\label{daA1}
d_a^{(A)}&=&6Q_{\rm AI}(1+2 w\cot w)~,\\\label{daD1} 
d_a^{(D)}&=&-{4 w\, Q_{\rm AI}\over\omega\tau_{\rm q}}~.
\eea
The nondiagonal symmetric term $h_s$ is due to the quadrupole mechanism:
\be
\label{hs1}
h_s=h_s^{(C)}={Q_{\rm AI}\over\wc\tau_{\rm q}}\, w\cot w~,
\ee
while the antisymmetric Hall part is generated by the photovoltaic mechanism:
\be\label{ha1}
h_a=h_a^{(D)}={Q_{\rm AS}\over\wc\tau_{\rm q}}(1+ w\cot w)~.
\ee

Let us now discuss the results (\ref{ds1})--(\ref{ha1}), obtained for
the linear regime in the microwave power:

\begin{itemize}
\item{the inelastic mechanism yields the dominant effect because of
$\tau_{\rm in}/\tau_{\rm q}\gg 1$: ${\rm (B)\gg (A),(C),(D)}$;}
\item{ ${\rm (C),(D)\gg (A)}$ for the case of $\wc\tau_{\rm
 q}\sim\omega\tau_{\rm q}\ll 1$ (strongly overlapping Landau levels). Note,
 however, that since $\delta$ [Eq.~(\ref{delta})] depends exponentially on
 $\omega_c$, a pronounced effect (from the practical point of view) is only
 possible at not too small magnetic fields. The effects of (A), (C), and
 (D) are then comparable in magnitude, differing only in $\ln \delta$;}
\item{although ${\rm (B)\gg (C),(D)}$, the oscillatory corrections to the
    Hall conductivity are entirely governed by the quadrupole (C) and
    photovoltaic (D) mechanisms;  }
\item{the quadrupole contribution $h_s$ violates Onsager symmetry:
    $\sigma_{ik}(B, s_-,s_+)\neq\sigma_{ki}(-B, s_+,s_-)$ [the effect
    of time inversion on the microwave field polarization can be expressed as
    $s_+\leftrightarrows s_-$, which is equivalent to
    $\omega\to-\omega$, see Eq.~(\ref{Epm})]. Although Onsager
    symmetry need not be satisfied in nonequilibrium systems, there
    exist not many examples of its violation. One of them, for a
    different system but under similar conditions, was found in
    Ref.~\onlinecite{antionsager}. Symmetry with respect to the parity
    transformation, $x\to-x$, must be satisfied: $\sigma_{ik}(B,
    s_-,s_+)=r_{ik}\sigma_{ik}(-B, s_+,s_-)$, where $r_{xx}=r_{yy}=1$,
    $r_{xy}=r_{yx}=-1$. It can be easily verified in our case since
    all parameters (\ref{w})--(\ref{QAS}) are unaffected by the
    simultaneous change $B\to-B$ and $s_+\leftrightarrows s_-$;}
\item{the quadrupole term $h_s$ and the anisotropic term in the
    diagonal photoresponse $d_a$ vanish in the case of circular
    polarization of the microwaves ($Q_{\rm AI}=0$).}
\end{itemize}

\section{Feedback effect: Saturation of the inelastic contribution
(SIC)} 
\label{s6}
\setcounter{equation}{0}

The results for $\hat\sigma_{\rm ph}\propto E_\omega^2$ obtained in
Sec.~\ref{s5}, are valid at not too high microwave power. The QBE in the form
(\ref{kinbot}) contains two sources of nonlinearity in $E_\omega^2$. At $Q\agt
1$, high-order terms in the expansion of $\hat{{\cal K}}_j$ and ${\cal
K}_\bot$ in powers of $E_\omega$ become important, leading to multiphoton
corrections to both $\phi_\bot$ and $\phi_j$, see Eqs.~(\ref{kinbot}) and
(\ref{phij}). The other source of nonlinearity, the ``feedback'' term ${\cal
K}_\bot\phi_\bot$ in Eq.~(\ref{kinbot}), strongly modifies the photoresponse
at much smaller microwave power, $Q_{\rm in}\sim 1$, where
\be
\label{Qin}
Q_{\rm in}\equiv\frac{\tau_{\rm in}}{2\tau_{\rm q}}Q~.
\ee
Since $\tau_{\rm in}\gg \tau_{\rm q}$, the multiphoton corrections can still
be neglected at $Q_{\rm in}\sim 1$.  As we show below in this section, at
$Q_{\rm in}\gg 1$ the single--photon feedback effect leads to a saturation of
the inelastic contribution and to a strong modification of all other
contributions to the photoconductivity.

\subsection{Feedback effect on the isotropic part of the distribution}
\label{ss61}

The feedback effect on the inelastic mechanism was studied in detail in
Ref.~\onlinecite{long}. Here, we reproduce the results\cite{long} within the
more general framework of Secs.~\ref{s3} and \ref{s4}. Using Eq.~(\ref{Kbot}),
we rewrite Eq.~(\ref{kinbot}) in the limit $Q\ll 1$ as
\be
\label{Q<1}
\left(\partial_t+\wc\partial_\varphi
+\frac{X^2}{\tau_{\rm q}}\right)\phi_\bot
+\frac{\langle\phi_\bot\rangle}{\tau_{\rm in}}
=\frac{2\partial_{ t_B}-\partial_t}{2\tau_{\rm q}}X^2~.
\ee
Averaging Eq.~(\ref{Q<1}) over $t$ and $\phi$, we obtain the equation for the
isotropic time-independent part
$\phi_\bot^{(B)}\equiv\langle\overline{\phi_\bot}\rangle$:
\bea
&&\left(\frac{\tau_{\rm q}}{\tau_{\rm in}}
+\left<\overline{X_\omega^2}\right>\right)\phi_\bot^{(B)}
\nonumber\\
&&=\partial_{ t_B}\left<\overline{X_\omega^2}\right>-
\left<\overline{
X_\omega^2\left(\phi_\bot-\phi_\bot^{(B)}\right)}\right>~.
\label{saturation}
\eea
In the linear regime with respect to the microwave power, i.e., for
$Q\ll\tau_{\rm q}/ \tau_{\rm in}$, we return to $\phi_\bot^{(B)}\propto Q$
given by Eq.~(\ref{phiB1}). At $Q\sim\tau_{\rm q}/\tau_{\rm in}\ll 1$ two
terms on the l.h.s. of Eq.~(\ref{saturation}) become of the same order of
magnitude, whereas the last term on the r.h.s. is still small and can be
neglected (corrections of higher order in $Q$, generated by this term, are
considered in Sec.~\ref{ss63}). We thus obtain for $Q\ll 1$ and arbitrary
order in $Q_{\rm in}$:
\be
\label{phiB}
\phi_\bot^{(B)}=\frac{Q_{\rm in}}{1+Q_{\rm in}}\omega\cot w~.
\ee
It follows that in the limit $Q_{\rm in}\gg 1$ we reach a saturation of
the inelastic contribution (SIC):
\be
\label{phiBsat}
\phi_\bot^{(B)}\to\frac{\partial_{ t_B}
\langle\overline{X_\omega^2}\rangle}{\langle\overline{X_\omega^2}\rangle}
=\omega\cot\frac{\pi\omega}{|\wc|}~.
\ee
The dc response in the stationary nonequilibrium state (\ref{phiB}) yields
\be
\label{dsB}
d_s^{(B)}=
\frac{4\,Q_{\rm in}}{1+Q_{\rm in}}\, 
w\cot w~,
\ee
reproducing Eq.~(15) of Ref.~\onlinecite{long}.

The saturation of the inelastic contribution at large $Q_{\rm in}$ has much in
common with the well-known effect of ``self-induced transparency'' for a
two-level system in a strong resonant electromagnetic field, which occurs due
to equalization of the population of the levels. Clearly, in the system we
consider here the complete transparency cannot be obtained since the electron
spectrum in our problem is continuous. Even in the SIC regime, absorption does
not tend to zero. Electrons still absorb microwave quanta and transmit the
absorbed power to the phonon bath: the dynamic equilibrium between these two
processes determines the effective electronic temperature $T_e$, see
Sec.~\ref{s9}. However, the saturation (\ref{phiBsat}) of the amplitude of the
oscillations in the isotropic part of the distribution function is
governed by the same feedback effect as the transparency in the two--level
system.

\subsection{Quadrupole, photovoltaic, and displacement contributions in the
SIC regime} 
\label{ss62}

In the nonlinear regime, the strong oscillations of $F_{00}(\ve)$ produced by
the inelastic mechanism essentially modify contributions (A), (C), and
(D). Since $\phi_\bot^{(C)}$ and $\phi_\bot^{(D)}$ are still small at $Q_{\rm
in}\sim 1$, we replace $\phi_\bot$ in the term $X^2\phi_\bot/\tau_{\rm q}$ on
the l.h.s.\ of Eq.~(\ref{Q<1}) by $\phi_\bot^{(B)}$. Solving
Eq.~(\ref{Q<1}) for the quadrupole and photovoltaic contributions then gives:
\bea
\label{phiCB}
\phi_\bot^{(CB)}&=&\partial_\varphi^{-1}\,
{\partial_{ t_B}-\phi_\bot^{(B)} \over \wc\tau_{\rm q}}
\left(
\overline{X_\omega^2}-\langle\overline{X_\omega^2}\rangle
\right)~,\\
\phi_\bot^{(DB)}&=&
2(\partial_t+\wc\partial_\varphi)^{-1}\nonumber\\
&\times&
\frac{\partial_{ t_B}-\partial_t/2-\phi_\bot^{(B)}}{\tau_{\rm q}}
\,X_\omega X_{\rm dc}~.
\label{phiDB}
\eea
Here and below we emphasize the influence of the inelastic mechanism by adding
the second superscript $B$. Substituting Eqs.~(\ref{phiCB}) and (\ref{phiDB})
into Eq.~(\ref{phij}) and using the resulting $\overline{\phi_j}$ in
Eq.~(\ref{Ovcur}) yields the photovoltaic and quadrupole contributions to the
Hall part of $\hat\sigma_{\rm ph}$ [Eq.~(\ref{sigmaph1})]:
\bea
\label{hsCB} 
&&
h_s^{(CB)}={Q_{\rm AI}\over\wc\tau_{\rm q}}\,
\frac{ w\cot w}{Q_{\rm in}+1}\,,\;\;\;
\\\label{haDB} &&
h_a^{(DB)}={Q_{\rm AS}\over\wc\tau_{\rm q}}
\left(1-\frac{Q_{\rm in}-1}{Q_{\rm in}+1} w\cot w\right)~.
\eea
The diagonal photovoltaic contributions $d_s^{(D)}$ and $d_a^{(D)}$, which
originate from the derivative $\partial_t$ in Eq.~(\ref{phiDB}), are not
affected by the oscillations in $F_{00}$ and remain the same as in the linear
regime, Eqs.~(\ref{dsD1}) and (\ref{daD1}).

The oscillations of $F_{00}$ also modify the displacement contribution to the
current. Similarly to Eqs.~(\ref{phiCB}) and (\ref{phiDB}) we substitute
$\phi_\bot^{(B)}$ for $\phi_\bot$ in Eq.~(\ref{phij}), which gives
\be
\label{phijAB}  
\overline{\phi_j^{(AB)}}=
\frac{\partial_{ t_B}-2\phi_\bot^{(B)}}{\wc\sqrt{\tau_{\rm q}\tau_{\rm tr}}}
\,6 X_{\rm dc}\overline{X_\omega^2}~.
\ee
Using Eq.~(\ref{phijAB}) in Eq.~(\ref{Ovcur}) we have
\bea
\label{dsAB}
&& d_s^{(AB)}=3Q\left(1-\frac{Q_{\rm in}-1}{Q_{\rm in}+1}\,2 w\cot w\right)~,
\\ &&
d_a^{(AB)}=6Q_{\rm AI}\left(1-\frac{Q_{\rm in}-1}{Q_{\rm in}+1}\,2 
w\cot w\right)~.\nonumber\\
\label{daAB}
\eea
We thus see that in the SIC regime, $Q_{\rm in}\gg 1$, the quadrupole term
$h_s^{(CB)}$ saturates as a function of the microwave power, whereas the
nondiagonal photovoltaic term $h_a^{(DB)}$ and the both displacement terms
$d_s^{(AB)}$ and $d_a^{(AB)}$ continue to grow linearly with increasing power
but are strongly modified as compared to the linear regime.

\subsection{Two--photon and double--frequency corrections to the isotropic
part of the distribution} 
\label{ss63}

For $Q_{\rm in}\gg 1$, the main (inelastic) contribution to the oscillatory
part of the photoconductivity is independent of the microwave power as long as
$Q$ is sufficiently small. Deviations from the ``plateau'' come from
contributions (A) and (D) (which, as shown in Sec.~\ref{ss62}, grow linearly
with $Q$ in the SIC regime), and also from corrections of order ${\cal O}(Q)$
to the $Q$ independent term $\phi_\bot^{(B)}$ [Eq.~(\ref{phiBsat})], which we
obtain below.

Equation (\ref{kinbot}) produces two types of corrections to
$\phi_\bot^{(B)}$ of linear order in $Q$.  One of them is related to
two--photon processes in $\hat{{\cal K}}_\bot$, which come from the
expansion of ${\cal K}_\bot$ to fourth order in ${\cal X}_\omega$,
yielding \bea \nonumber &&\phi_\bot^{(B)}=\frac{\partial_{ t_B}
\langle\overline{X_\omega^2-X_\omega^4}\rangle}{\tau_{\rm q}\tau_{\rm
in}^{-1} +\langle\overline{X_\omega^2-X_\omega^4}\rangle}\\
&&=\omega\cot w\,\frac{Q_{\rm in}}{Q_{\rm in}+1} \left(1-{3\over 4} Q
\frac{Q_{\rm in}+2}{Q_{\rm in}+1}\right)~.  \label{phiB2photon} \eea
The other is associated with single--photon excitation of the second
temporal harmonic of the distribution function. This correction
originates from the term $\partial_t X_\omega^2$ in Eq.~(\ref{Q<1})
(which has played no role so far) and to leading order in $X_\omega^2$
reads: \be \label{phiB2w} \phi_\bot^{(2\omega)}=C-\frac{(2\tau_{\rm
q})^{-1}} {\partial_t+\omega_c\partial_\varphi}\,
\partial_t\left(X_\omega^2-\overline{X_\omega^2}\right)~, \ee where
\bea X_\omega^2-\overline{X_\omega^2}&=& {1\over 2} \sin^2 w \left[\,
{\cal E}_+^2\cos(2\omega t+2\varphi)\right.\nonumber\\ &+& \left.{\cal
E}_-^2 \cos(2\omega t-2\varphi)+2{\cal E}_+{\cal E}_-\cos 2\omega t\,
\right]~.\nonumber\\ \label{X^2rest} \eea The constant of integration
$C$ in Eq.~(\ref{phiB2w}) should be found from the periodic boundary
condition for the function $\phi_\bot^{(2\omega)}(\varphi, t)$.
Averaging Eq.~(\ref{Q<1}) over $\varphi$ and $t$, we get the boundary
condition in the form $\tau_{\rm in}\langle \overline{X_\omega^2
\phi_\bot^{(2\omega)}} \rangle=-\tau_{\rm q}\langle\overline{
\phi_\bot^{(2\omega)}}\rangle$, which gives $C=\tau_{\rm
q}^{-1}Q_{2\omega}Q_{\rm in}/(1+Q_{\rm in})$, where \bea
&&Q_{2\omega}=\frac{\sin^4 w}{8 Q}\left(4{\cal E}_+^2 {\cal E}_-^2
+\frac{\omega}{\omega+\omega_c}{\cal E}_+^4
+\frac{\omega}{\omega-\omega_c}{\cal E}_-^4\right)~.\nonumber\\
\label{Q2w} \eea Only the constant $C$ contributes to the
double-frequency correction to the dc current at first order in $Q$.
Summing up the contributions (\ref{phiB2photon}) and (\ref{phiB2w}) to
$\phi_\bot^{(B)}$ we obtain $d_s^{(B)}=2 t_B\phi_\bot^{(B)}$ with the
linear-in-$Q$ terms included: \bea &&d_s^{(B)}\simeq\frac{4 w\, Q_{\rm
in}}{1+Q_{\rm in}}\, \left[\,\cot w\left(1-{3\over4}Q\,\frac{Q_{\rm
in}+2}{Q_{\rm in}+1}\right) +\frac{Q_{2\omega}}{\omega\tau_{\rm
q}}\,\right]~.\nonumber\\ \label{dsBfull} \eea
Equation~(\ref{dsBfull}) tells us that for strongly overlapping Landau
levels the double-frequency correction is much stronger than the one
coming from the two-photon processes, except for the close vicinity of
the cyclotron-resonance harmonics where the two-photon correction
 exhibits singular behavior.

\subsection{Discussion of the results}
\label{ss64}

Equations (\ref{hsCB}), (\ref{haDB}), (\ref{dsAB}), (\ref{daAB}), and
(\ref{dsBfull}), together with Eqs.~(\ref{dsD1}) and (\ref{daD1}),
describe the photoconductivity $\hat{\sigma}_{\rm ph}$
[Eq.~(\ref{sigmaph1})] to arbitrary order in $Q_{\rm in}=\tau_{\rm
in}Q/2\tau_{\rm q}$ and to first order in $Q$. In the limit $Q_{\rm
in}\ll 1$, Eqs.~(\ref{ds1})--(\ref{ha1}) are reproduced.  At $Q_{\rm
in}\agt 1$, the evolution of the photoresponse with increasing power
can be summarized as following:

\begin{itemize}
\item{Growth of the
leading (inelastic) term $d_s^{(B)}$ saturates due to the feedback effect,
Eq.~(\ref{dsB}), and so does the quadrupole term $h_s^{(CB)}$,
Eq.~(\ref{hsCB}). }
\item{Remarkably, the parts proportional to $ w\cot w$
of the displacement contributions $d_s^{(AB)}$ and $d_a^{(AB)}$, as well as
that of the photovoltaic contribution $h_a^{(DB)}$, change sign at $Q_{\rm
in}=1$, and the linear growth is recovered with  opposite sign at higher power.}
\item{While at $Q_{\rm in}\ll 1$ the
inelastic contribution is the only one that is $T$ dependent, in the crossover
region, $Q_{\rm in}\sim 1$, all contributions, except for the diagonal
photovoltaic terms $d_s^{(D)}$ and $d_a^{(D)}$, change strongly as a function
of $T$.}
\item{ However, apart from the small quadrupole correction $h_s^{(CB)}$, the
photoconductivity in the limit $Q_{\rm in}\gg 1$ becomes independent of
$\tau_{\rm in}$ and, hence, of temperature:}
\end{itemize}
\bea
\label{dsBsat}
d_s^{(B)}&=& w\cot w(4-3 Q)+4 w Q_{2\omega}/\omega\tau_{\rm q}~,\\
\label{dsABsat} 
d_s^{(AB)}&=&3Q\left(1-2 w\cot w\right)~,\\\label{daABsat} 
d_a^{(AB)}&=&6Q_{\rm AI}(1-\,2 w\cot w)~,\\\label{dsDsat} 
d_s^{(D)}&=& w\,Q_{\rm S}/\omega\tau_{\rm q}~,\\\label{daDsat} 
d_a^{(D)}&=&-4 w\, Q_{\rm AI}/\omega\tau_{\rm q}~,\\\label{hsCBsat} 
h_s^{(CB)}&=&(Q_{\rm AI}/Q_{\rm in})w\cot w/\omega_c\tau_{\rm q}~,\\
\label{haDBsat}
h_a^{(DB)}&=&Q_{\rm AS}(1- w\cot w )/\wc\tau_{\rm q}~.
\eea
Equations~(\ref{dsBsat})--(\ref{haDBsat}) predict that there are two sources
of nonlinearities which will essentially modify the plateau behavior in the
SIC regime with increasing microwave power: the multiphoton processes and the
``spreading'' of the feedback effect over higher temporal and angular
harmonics. The multiphoton effects become strong at $Q\agt 1$, whereas the
feedback-related excitation of distant harmonics becomes strong already at much
lower power, namely at $Q\agt \omega_c\tau_{\rm q}$. It is important that the
strong deviations from the plateau behavior take place when $Q$ is already
much larger than $\tau_{\rm q}/\tau_{\rm in}$ and, therefore, inelastic
scattering can be completely neglected. In Sec.~\ref{s7} we discuss the regime
of ultrahigh power in more detail.
 
\section{Ultrahigh power: Feedback vs multiphoton effects}
\label{s7}
\setcounter{equation}{0}

\subsection{Classification of contributions to $\hat{\sigma}_{\rm ph}$ at
  ultrahigh power } 
\label{ss71}

In the SIC regime considered in Sec.~\ref{s6}, strong oscillations in
isotropic part of the distribution function essentially modified all
contributions to the dc current (A)--(D).  At ultrahigh power,
contributions involving high angular and temporal harmonics of the
distribution function become important, so that our previous (A)--(D)
classification, developed in Sec.~\ref{s3} for the linear
$\hat{\sigma}_{\rm ph}$, should be generalized.  Despite the excitation of the high
harmonics, at ultrahigh power it is still possible to distinguish the
generalized displacement (A), photovoltaic (D), and non-photovoltaic
 contributions.  In the linear regime with respect to the dc field
$E_{\rm dc}$, the latter enters the part of the kernel $\hat{\cal
  K}_\bot$ in the photovoltaic term only. In all other terms, $E_{\rm
  dc}$ belongs to $\hat{\cal K}_j$. Further, in the displacement term
all effects of the microwave and dc fields on even angular harmonics
are absent. We now proceed to calculate
various contributions to $\hat{\sigma}_{\rm ph}$ in the limit of
ultrahigh power, where the inelastic scattering is of no importance.

The displacement contribution is readily calculated to arbitrary order in
microwave power by putting $\phi_\bot=0$ in Eq.~(\ref{phij}) for $\phi_j$ and
extracting the part linear in $E_{\rm dc}$:
\bea
\phi_{j}^{(A)}
&=&\frac{e v_F}{\wc^2\tau_{\rm tr}}\,{\rm Re}\,\left(E_{\rm dc}^{+}\, 
e^{- i\varphi}\right)\nonumber\\
&\times&\partial_{t_B}t_B
\frac{1-3 X_\omega^2}{(1+X_\omega^2)^3}~,
\label{phijA}
\eea
where $E_{\rm dc}^{+}=(\bE_{\rm dc})_x+i(\bE_{\rm dc})_y$ and we used
the explicit form of $X_{\rm dc}$ given by Eq.~(\ref{Xdc}). As for the
effect of the ac and dc fields on even angular harmonics,
Eq.~(\ref{phij}) yields:
\bea
\label{phij0}
&&\phi_{j}^{(0)}=\frac{2\phi_{\bot }^{(0)} X_{\rm dc}}{\wc\sqrt{\tau_{\rm q}
\tau_{\rm tr}}}\,
\frac{1-3 X_\omega^2}{(1+X_\omega^2)^3}~,
\\\label{phij1}
&&\phi_{j}^{(D)}
=\frac{2\phi_{\bot}^{(D)}}{\wc\sqrt{\tau_{\rm q}\tau_{\rm tr}}}\,
\frac{X_\omega}{(1+X_\omega^2)^2}~.
\eea
Here $\phi_{\bot }^{(0)}$ is the solution of Eq.~(\ref{kinbot}) that is
independent of $E_{\rm dc}$ (at small power, $\phi_{\bot }^{(0)}$ is the sum
of the inelastic and quadrupole terms) and $\phi_{\bot}^{(D)}$ is the
linear-in-$E_{\rm dc}$ (photovoltaic) solution:
\bea
\nonumber
&&\left(\partial_t+\wc\partial_\varphi+{1\over\tau_{\rm q}}
\frac{X_\omega^2}{1+X_\omega^2}\right)\phi_{\bot}^{(0)}
\\\label{kinbot1}
&&\quad=\frac{\partial_{t_B}-\partial_t/2}{\tau_{\rm q}}
\frac{X_\omega^2}{1+X_\omega^2}~,\\\nonumber
&&\left(\partial_t+\wc\partial_\varphi+{1\over\tau_{\rm q}}
\frac{X_\omega^2}{1+X_\omega^2}\right)\phi_{\bot}^{(D)}
\\\label{kinbot2}
&&\quad=
\frac{\partial_{t_B}-\partial_t/2-\phi_{\bot}^{(0)}}{\tau_{\rm q}}\,
\frac{2 X_{\rm dc} X_\omega}{(1+X_\omega^2)^2}~.
\eea
In Eq.~(\ref{kinbot1}), we omitted the term which describes inelastic
scattering, since in the following we consider the limit $Q\gg \tau_{\rm
q}/\tau_{\rm in}$. In this limit, the inelastic scattering is irrelevant, as
demonstrated in Sec.~\ref{ss64}. For simplicity, we focus here on the case of
circularly polarized microwaves. Linear polarization is briefly discussed in
the end of Sec.~\ref{ss73}.

\subsection{Circular polarization: General solution}
\label{ss72}

For a circularly polarized microwave field, solution of Eqs.~(\ref{kinbot1})
and (\ref{kinbot2}) greatly simplifies. For definiteness, let us consider 
passive circular polarization by putting $s_+=1$, $s_-=0$ in Eq.~(\ref{X}), so
that
\bea
\label{X+}
&&X_\omega=\sqrt{Q}\cos  x\,,\qquad  x\equiv\varphi+\omega t~,\\\label{QX+}
&&Q=\frac{2\tau_{\rm q}}{\tau_{\rm tr}}
\left[\frac{e E_\omega v_F}{\omega(\wc+\omega)}\right]^2
\sin^2 w~.
\eea
It is convenient to introduce new functions $g_0(x)$ and $g_1(x)$ by casting
$\phi_{\bot}^{(0)}(x)$ and $\phi_{\bot}^{(D)}(x,\varphi)$ in the form
\bea
\label{g0}
\phi_{\bot}^{(0)}&=&2 t_B^{-1}\,g_0(x)~,\\
\phi_{\bot}^{(D)}&=&-\frac{e v_F}{\wc}\sqrt{\frac{\tau_{\rm q}}{\tau_{\rm tr}}}
\nonumber\\ 
&\times&
2 \,{\rm Re}\left[ \,E_{\rm dc}^{(+)}\,g_1(x)\, \exp(- i\varphi)\,\right]~.
\label{g1}
\eea
The partial differential equations (\ref{kinbot1}) and (\ref{kinbot2})
transform then into ordinary differential equations for $g_{0,1}(x)$: 
\be
g_N^{\prime}(x)+ \left[\alpha\, p(x)-i\beta \, \delta_{N 1}\right] g_N(x)
=\alpha \Pi_N(x)~, 
\label{kineqg} 
\ee 
where $N=0$ or 1, the constants $\alpha$ and $\beta$ are given by 
\be 
\alpha=1/(\omega+\wc)\tau_{\rm q}~,\quad \beta=\wc\tau_{\rm q}\alpha~, 
\label{alpha} 
\ee 
and the functions $p(x)$ and $\Pi_N(x)$ are written in terms of $X_\omega$ as
\bea 
\label{p} 
&& p(x)=\frac{X_\omega^2}{1+X_\omega^2}~,\\
\label{Pi0} 
&&\Pi_0(x)=w(\cot w+\tan x)\frac{X_\omega^2}{(1+X_\omega^2)^2}~,\\\nonumber &&
\Pi_1(x)=\left[\,1-2 g_0(x)\,\right]\, \frac{X_\omega}{(\,1+X_\omega^2\,)^2}\\
&& \phantom{a}\qquad +w(\,\cot w+\tan x\,)X_\omega
\frac{1-3X_\omega^2}{(\,1+X_\omega^2\,)^3}~.  \nonumber\\ 
\label{Pi1} 
\eea
Integration of Eq.~(\ref{kineqg}) with a periodic boundary condition
$g_N(0)=g_N(2\pi)$ yields 
\bea 
&& g_N(x)\!=\!{\cal G}_N(x)+\exp[-S_N(x)\,]\,
\frac{{\cal G}_N(2\pi)}{1\!-\!\exp[- S_N(2\pi)]}~,\nonumber\\
\label{solution}\\ &&{\cal G}_N(x)=\alpha\int\limits_0^x\! d y\,
\Pi_N(y)\exp[\,S_N(y)-S_N(x)\,]~,\nonumber\\ \label{GN} \\
\label{SN}
&&S_N(x)=-i\,\beta\,x\,\delta_{N 1}+\alpha\,\int\limits_0^x \! d y \,p(y)~.
\eea 

Note that for the case of circular polarization the anisotropic contributions
to the photoconductivity tensor $\hat{\sigma}_{\rm ph}$ [Eq.~(\ref{sigmaph1})]
vanish, $d_a=h_s=0$, so that $\hat{\sigma}_{\rm ph}$ reduces to 
\bea 
\hat{\sigma}_{\rm ph}={2\sigma^{\rm D}_{xx}}
\left(\!\begin{array}{cc}1+2\delta^2(1-d_s)&-\wc\tau_{\rm tr}-2\delta^2 h_a\\
\wc\tau_{\rm tr}+2\delta^2
h_a&1+2\delta^2(1-d_s)\end{array}\!\right)~.\nonumber\\ 
\label{cirph} 
\eea 
By using Eqs.~(\ref{Ovcur}), (\ref{phijA})--(\ref{phij1}), (\ref{g0}), and
(\ref{g1}), the terms $d_s$ and $h_a$ in Eq.~(\ref{cirph}) are expressed as
\be 
d_s=d_s^{(A)}+d_s^{(0)}+d_s^{(D)}~, 
\label{ds} 
\ee 
where 
\bea
&&d_s^{(A)}=1-(1+2 w \cot w\,Q\,\partial_Q) \left\langle\frac{1-3X_\omega^2}
{(\,1+X_\omega^2\,)^3}\right\rangle_x~,\nonumber\\ 
\label{dsa}
\\ 
&&d_s^{(0)}=4\left\langle g_0(x) \frac{1-3X_\omega^2}{(\,1+X_\omega^2\,)^3}
\right\rangle_x~,
\label{ds0}
\\ &&d_s^{(D)}=4\,{\rm Re}\,\left\langle g_1(x)
\frac{X_\omega}{(\,1+X_\omega^2\,)^2}\right\rangle_x~, 
\label{dsd} 
\eea 
and
\be 
h_a=-4\,{\rm Im}\,\left\langle g_1(x)
\frac{X_\omega}{(\,1+X_\omega^2\,)^2}\right\rangle_x~. 
\label{ha} 
\ee 
The brackets $\left<\dots\right>_x$ denote averaging over $x$.



\subsection{Photoresponse at ultrahigh power}
\label{ss73}

Being represented in the form of Eq.~(\ref{dsa}), the displacement
contribution is easily calculated,
\be
d_s^{(A)}
=1+\frac{Q/2-1}{(1+Q)^{5/2}}+2 w \cot w\,Q\frac{3-Q/4}{(1+Q)^{7/2}}~,
\label{dsA}
\ee
reproducing Eq.~(6.8) of Ref.~\onlinecite{VA}. Unlike $d_s^{(A)}$, the terms
$d_s^{(0)}$, $d_s^{(D)}$, and $h_a$, which are expressed through the
functions $g_0(x)$ and $g_1(x)$, cannot be written in a closed analytical form
for arbitrary $\alpha$, $\beta$, and $Q$. When discussing $\hat{\sigma}_{\rm
ph}$ at ultrahigh power in various limiting cases, it is useful to represent
$d_s$ and $h_a$ as 
\bea
\nonumber
d_s&=&w \cot w \,{\rm F_{1d}}(Q,\alpha,\beta)\\\label{dsF}
&+&w\, {\rm F_{2d}}(Q,\alpha,\beta)
+{\rm F_{3d}}(Q,\alpha,\beta)~,\\\nonumber
h_a&=&w \cot w \,{\rm F_{1h}}(Q,\alpha,\beta)\\\label{haF}
&+&w\,{\rm F_{2h}}(Q,\alpha,\beta)
+{\rm F_{3h}}(Q,\alpha,\beta)~,
\eea
where the different terms are in one-to-one correspondence with the terms in
the functions $\Pi_N(x)$ on the r.h.s.\ of Eq.~(\ref{kineqg}) proportional to
$w \cot w$, $w$, and independent of $w$ (except for the dependence on $\omega$
in $X_\omega$). 

\begin{figure}[ht]
\centerline{ 
\includegraphics[width=0.95\columnwidth]{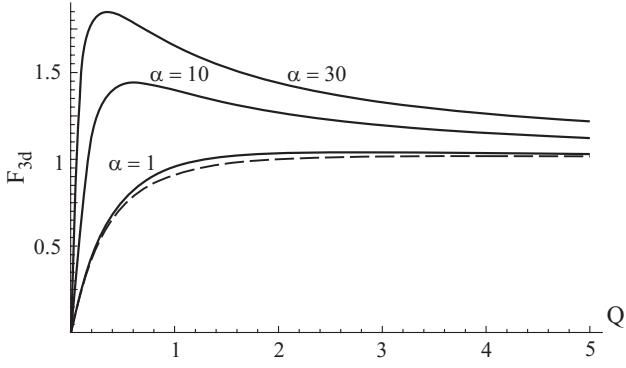}}
\caption{Evolution of the contribution $F_{3d}(Q)$, Eq.~(\ref{dsF}),
  to the photoconductivity (\ref{cirph}) with increasing effective
  microwave power $Q$, Eq.~(\ref{QX+}), for $\alpha=1,\,10,\,30$, and
  $\beta=0.3$ [Eq.~(\ref{alpha})].  Dashed curve: the displacement
  contribution to $F_{3d}(Q)$, given by the first two terms in
  Eq.~(\ref{dsA}).}
 \label{fF3d}
 \end{figure}

\begin{figure}[ht]
\centerline{ 
\includegraphics[width=0.95\columnwidth]{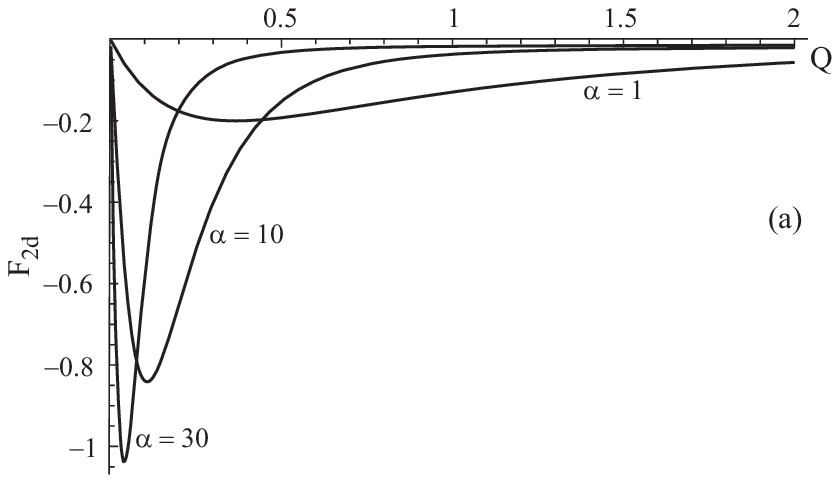}
}
\centerline{ 
\includegraphics[width=0.95\columnwidth]{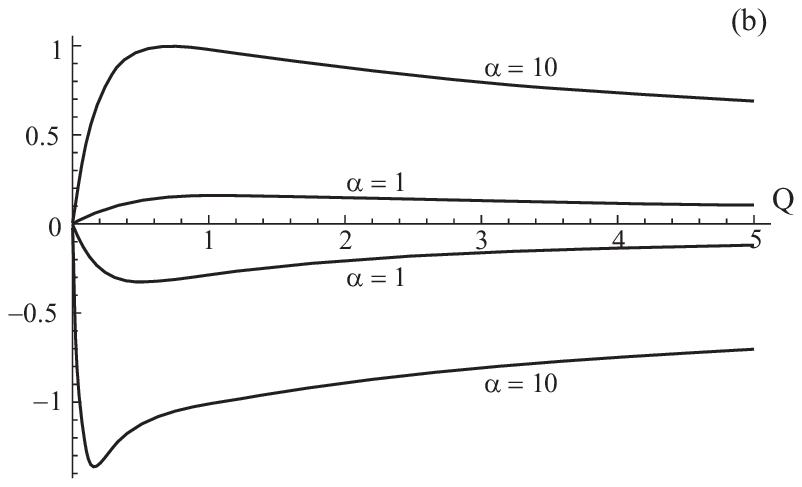}
}
\caption{(a) Contribution $F_{2d}(Q)$, Eq.~(\ref{dsF}), to the
  photoconductivity (\ref{cirph}) for $\alpha=1,\,10,\,30$, and
   $\beta=0.3$; (b) ``Nonphotovoltaic'' (positive) and photovoltaic
   (negative) contributions to $F_{2d}(Q)$ for $\alpha=1,\,10$. }
\label{fF2d}
\end{figure}

\begin{figure}[ht]
\centerline{ 
\includegraphics[width=0.95\columnwidth]{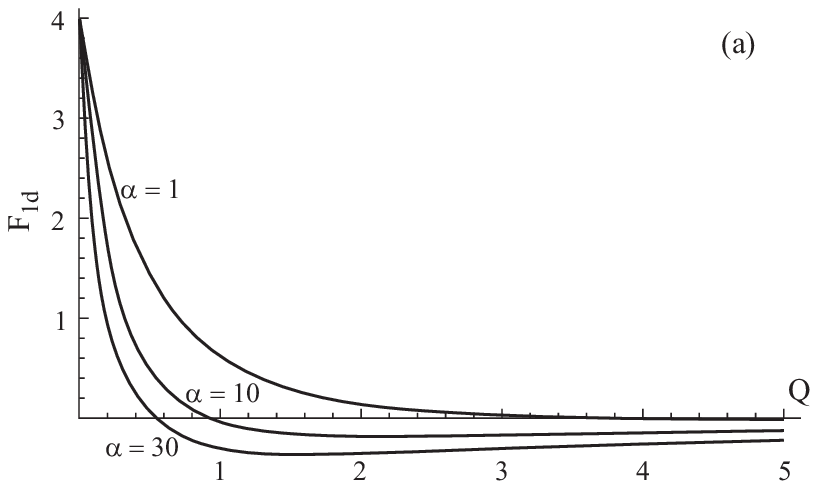}
}
\centerline{ 
\includegraphics[width=0.95\columnwidth]{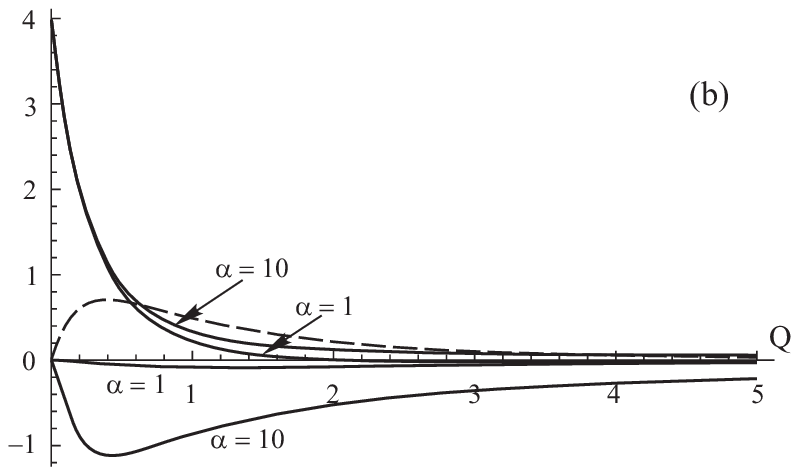}
}
\caption{(a) Contribution $F_{1d}(Q)$, Eq.~(\ref{dsF}), to the
  photoconductivity (\ref{cirph}) for $\alpha=1,\,10,\,30$, and
  $\beta=0.3$; (b) Contributions to $F_{1d}(Q)$ [Eq.~(\ref{ds})] for
  $\alpha=1,\,10$: Dashed line: Displacement contribution [third term
  in Eq.~(\ref{dsA})]; solid line in the upper half-plane:
  nonphotovoltaic contribution [second term in Eq.~(\ref{ds})]; solid
  line in the lower half-plane: photovoltaic contribution [third term
  in Eq.~(\ref{ds})].  }
\label{fF1d}
\end{figure}

\begin{figure}[ht]
\centerline{ 
\includegraphics[width=0.95\columnwidth]{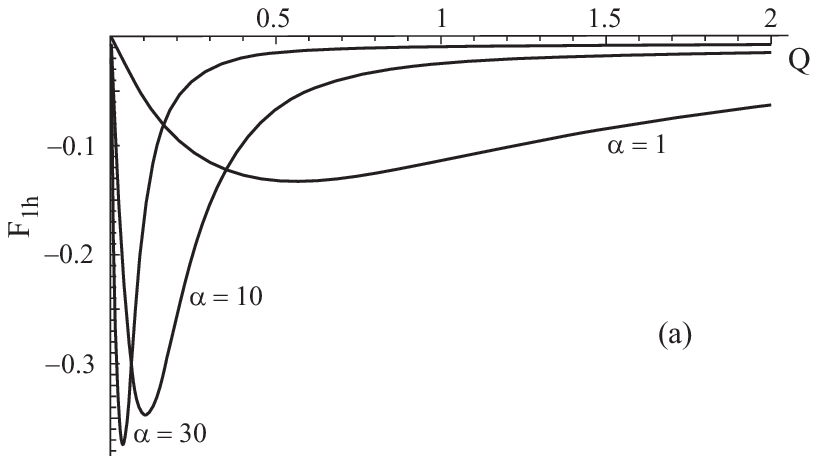}
}
\centerline{ 
\includegraphics[width=0.95\columnwidth]{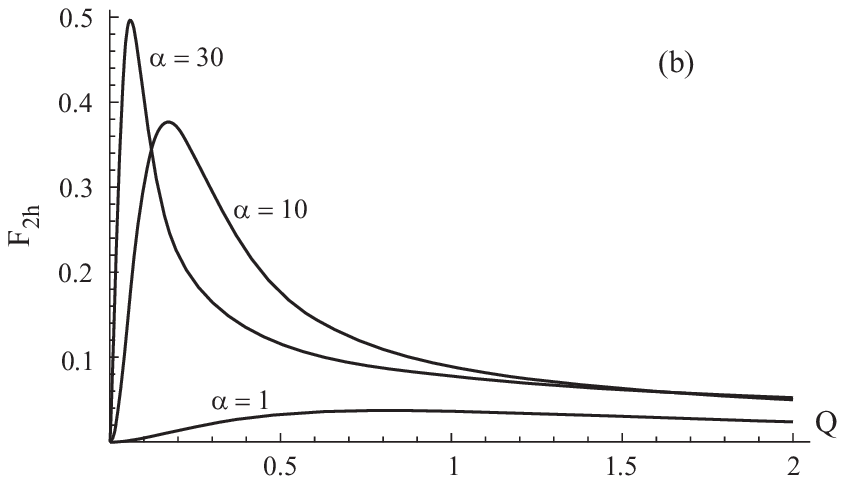}
}
\centerline{ 
\includegraphics[width=0.95\columnwidth]{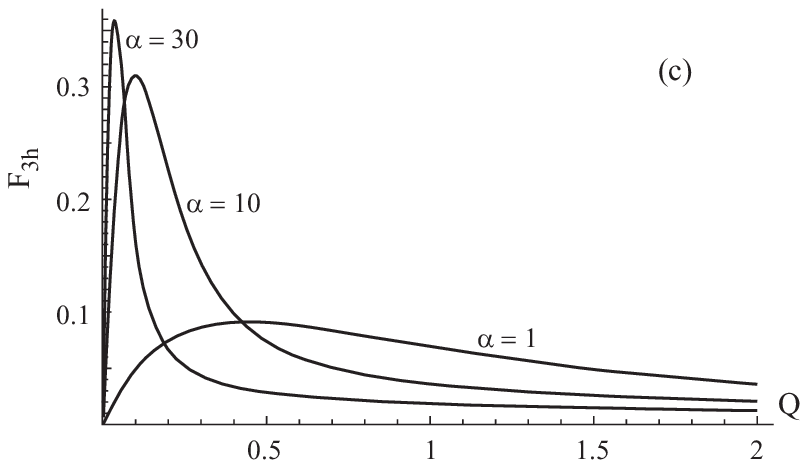}
}
\caption{ Hall contributions (\ref{haF}) to the photoconductivity 
(\ref{cirph}) for $\alpha=1,\,10,\,30$, and $\beta=0.3$:
(a) $F_{1h}(Q)$; (b) $F_{2h}(Q)$; (c)~$F_{3h}(Q)$.}
\label{fFh}
\end{figure}

The results of a numerical calculation of $d_s$ and $h_a$
[Eqs.~(\ref{dsa})--(\ref{ha})] are illustrated in Figs.~\ref{fF3d}--\ref{fFh},
where ${\rm F}_{nd}$ and ${\rm F}_{nh}$ (with $n=1,\,2,\,3$) are drawn as a
function of $Q$ for $\beta=0.3$ and several values of $\alpha=1,10,30$. Since
$\tau_{\rm in}$ is assumed to be infinite throughout Sec.~\ref{s7}, the scale
of $Q\sim\tau_{\rm q}/\tau_{\rm in}$, below which the system exhibits the
linear response with respect to the microwave power, vanishes. As a result,
the function ${\rm F}_{1d}$ is nonzero (${\rm F}_{1d}=4$) at $Q=0$ in
Fig.~\ref{fF1d}. With the exception of ${\rm F}_{3d}$, all other functions
${\rm F}_{nd}$ and ${\rm F}_{nh}$ show nonmonotonic behavior with increasing
$Q$ and decay to zero in the limit of large $Q$. The function ${\rm F}_{3d}\to
1$ for $Q\to\infty$, thus eliminating, as can be seen from Eq.~(\ref{cirph}),
the quantum correction $2\delta^2$ to the dissipative part of the classical
Drude conductivity. It follows that all quantum effects in $\hat{\sigma}_{\rm
ph}$ are destroyed and the classical Drude conductivity is restored in the
limit of high power.

For $\alpha\gg 1$ (which is the case for strongly overlapping Landau levels,
unless $\omega/\omega_c$ is too large), the displacement contribution
(\ref{dsA}) is clearly insufficient to correctly describe the behavior of
$d_s$ as a function of $Q$. This can be seen, e.g., in Fig.~\ref{fF3d}, where
the displacement contribution to ${\rm F}_{3d}$ is shown by the dashed line.
A much faster behavior of ${\rm F}_{3d}$ at small $Q$ is due to the
photovoltaic contribution. Indeed, the expansion of ${\rm F}_{3d}$ in powers
of $Q$ reads:
\be
{\rm F}_{3d}
\simeq 3 Q+{(\alpha Q)^2\over 2}\frac{1+3\beta^2}{(1-\beta^2)^2}~,
\qquad Q\ll{1\over\alpha}~,
\label{F3d}
\ee
where the linear term, due to the displacement mechanism, crosses over at
$Q\sim \alpha^{-2}$ into the faster quadratic term related to the photovoltaic
mechanism. At $Q\sim \alpha^{-1}$, ${\rm F}_{3d}$ reaches a maximum value of
order unity. In the limit of large $Q$, ${\rm F}_{3d}$ approaches unity from
above.

The displacement contribution to the function ${\rm F}_{2d}$, shown in
Fig.~\ref{fF2d}(a), is exactly zero, so that ${\rm F}_{2d}$ is entirely due to
the contribution of $d_s^{(0)}$ and $d_s^{(D)}$. At $Q\ll\alpha^{-1}$, ${\rm
F}_{2d}$ behaves linearly with $Q$:
\be
{\rm F}_{2d}\simeq {\alpha Q\over 2}- {2\alpha Q\over 1-\beta^2}~.
\label{F2d}
\ee
The first term in Eq.~(\ref{F2d}) comes from $d_s^{(0)}$ and is associated
with excitation of the double-frequency harmonics $F_{22}$ and $F_{-2,-2}$
[Eqs.~(\ref{phiB2w}) and (\ref{dsBsat})]. The second, photovoltaic term, which
is due to excitation of $F_{21}$, $F_{01}$, $F_{-2,-1}$, and $F_{0,-1}$, has a
larger slope with opposite sign [Eq.~(\ref{dsDsat})]. Their behavior at
arbitrary $Q$ is illustrated in Fig.~\ref{fF2d}(b): one can see that each term
decreases slowly with increasing $Q$ for $Q\gg\alpha^{-1}$, but they strongly
compensate each other, leading to a much faster decay of ${\rm F}_{2d}$.

The function ${\rm F}_{1d}$ is shown in Fig~\ref{fF1d}(a). It is
finite at $Q=0$ and therefore is much larger than both ${\rm F}_{2d}$
and ${\rm F}_{3d}$ at $Q\ll\alpha^{-1}$. The expansion of ${\rm
F}_{1d}$ in powers of $Q$ reads:
\be
{\rm F}_{1d}
\simeq 4- 9Q -{(\alpha Q)^2\over 2}\frac{1+3\beta^2}{(1-\beta^2)^2}~.
\label{F1d}
\ee
The nonzero value of ${\rm F}_{1d}$ at $Q=0$ is due to the saturated
inelastic contribution $d_s^{(B)}$ [Eq.~(\ref{dsBsat})], whereas the
linear term comes from both the two-photon correction to the inelastic
contribution [Eq.~(\ref{dsBsat})] and from the displacement
contribution [Eq.~(\ref{dsABsat})]. The two contributions related to
$d_s^{(0)}$ and $d_s^{(A)}$ are presented for arbitrary $Q$ in
Fig.~\ref{fF1d}(b). Similarly to ${\rm F}_{3d}$, the photovoltaic term
grows quadratically at $Q\ll \alpha^{-1}$. As seen from
Fig.~\ref{fF1d}(b), for $\alpha\gg 1$ it dominates at large $Q$, so
that ${\rm F}_{1d}$ changes sign, reaches negative values of order
unity, and approaches zero in the limit $Q\to\infty$ from below.

The functions ${\rm F}_{nh}$ contributing to the Hall part of
$\hat{\sigma}_{\rm ph}$ are shown in Fig.~\ref{fFh}. All three are related to
the photovoltaic mechanism and come from the imaginary part of $g_1(x)$ in
Eq.~(\ref{ha}). At $Q\ll\alpha^{-1}$ they grow with increasing microwave
power,
\bea
\label{F1h}
&&{\rm F}_{1h}\simeq-\frac{2\beta}{1-\beta^2}\alpha Q~,\\\label{F2h}  
&&{\rm F}_{2h}\simeq\frac{2\beta}{(1-\beta^2)^2}(\alpha Q)^2~,\\\label{F3h}
&&{\rm F}_{3h}\simeq\frac{2\beta}{1-\beta^2}\alpha Q~,
\eea
and decay to zero in the limit of large $Q$. The maximum values of $|{\rm
F}_{nh}|$, which are reached at $Q\sim\alpha^{-1}$, are of order unity (for
$\beta\sim 1$), similarly to the behavior of ${\rm F}_{nd}$. It follows that
at $Q\sim\alpha^{-1}$ the Hall part $h_a$ of the photoresponse becomes as
strong as the diagonal part $d_s$.

The behavior of $d_s$ and $h_a$ in the limit of large $Q$ can be found by
observing that the main contribution to the integrals (\ref{dsa})--(\ref{ha})
over $x$ comes then from a close vicinity (of width $Q^{-1/2}$) of points
$x=\pi/2$ and $x=3\pi/2$. Around these points the multiphoton processes, which
sum up to produce the factors $(1+X_\omega^2)^{-1}$ in
Eqs.~(\ref{dsa})--(\ref{ha}), are strongly suppressed. For $Q\gg\alpha^2$, the
expansion of the factors $\exp (\pm S_N)$ around unity in Eq.~(\ref{GN})
yields the asymptotic behavior of the functions ${\rm F}_{nd}$ and ${\rm
F}_{nh}$:
\bea
\label{F1d>} 
&&{\rm F}_{1d}\simeq -\frac{7\alpha^2}{16 Q^{3/2}}~,\\
\label{F3d>} 
&&{\rm F}_{3d}\simeq 1+\frac{\alpha^2}{8 Q^{3/2}}~,\\
\label{F1h>}
&&{\rm F}_{1h}\simeq -\frac{3\beta\alpha}{4 Q^{3/2}}~,\\
\label{F3h>}
&&{\rm F}_{3h}\simeq \frac{\beta\alpha}{2 Q^{3/2}}~.
\eea 
Note that the displacement mechanism gives terms of order $Q^{-3/2}$
in ${\rm F}_{1d}$ and ${\rm F}_{3d}$, which are much smaller than the
terms of order $\alpha^2 Q^{-3/2}$ in Eqs.~(\ref{F1d>}) and
(\ref{F3d>}) for $\alpha\gg 1$.

In the intermediate interval $\alpha^{-1}\ll Q\ll \alpha^2$, where all the
functions ${\rm F}_{nd}$ and ${\rm F}_{nh}$, $n=1,2,3$, fall off with increasing $Q$ as a
power law, the function $g_N(x)$ can be approximated as $g_N(x)\simeq {\cal
G}_N(x)\simeq \Pi_N(x)/p(x)$ everywhere except for a vicinity of the singular
points $x=\pi/2$ and $x=3\pi/2$. The singular behavior around these points
requires special care for the function ${\rm F}_{2d}$, contributions to which
coming from $d_s^{(0)}$ and $d^{(D)}$ only weakly depend on $Q$ and strongly
compensate each other, as illustrated in Fig.~\ref{fF2d}(b). Consider the part
${\rm F}^{(0)}_{2d}$ related to $d_s^{(0)}$:
\be
{\rm F}^{(0)}_{2d}\simeq {2\over\pi}\int\limits_0^{2\pi}\!dx\,
\zeta(x)\frac{1-3Q\cos^2 x}{(1+Q\cos^2 x)^3}~,
\label{F02d}
\ee
where
\bea
\nonumber \zeta(x)&=&\alpha Q\int\limits_0^{x}\!d y \frac{\sin y \cos
y}{(1+Q\cos^2 y)^2}\\\label{gw}
&\times&\exp\left(-\int\limits_y^x\!dz\frac{\alpha Q \cos^2 z}{1+Q\cos^2 z}
\right)~. 
\eea 
In Eq.~(\ref{F02d}) we neglected the exponentially small (for
$Q\gg\alpha^{-1}$) boundary term in the solution (\ref{solution}). For
$Q\gg\alpha^{-1}$ the main contribution to ${\rm F}^{(0)}_{2d}$ given by
Eq.~(\ref{F02d}) comes from the singularities at $x=\pi/2,3\pi/2$. For
$Q\ll\alpha^2$ one can neglect $Q$ in all the denominators in
Eqs.~(\ref{F02d}) and (\ref{gw}), after which ${\rm F}^{(0)}_{2d}$ is
represented as
\bea 
\nonumber 
{\rm F}^{(0)}_{2d}&\simeq &{2\alpha
Q\over\pi}\int\limits_{-\infty}^{\infty}\!d x_1
\int\limits_{-\infty}^{\infty}\!d x_2\;x_1\, {\rm sgn}(x_1-x_2)\\
\label{asymptote} &\times&\exp\left(-\alpha Q{|x_2^3-x_1^3|\over
3}\right)={4\over\sqrt{3}}~.  
\eea 
The result does not depend on $Q$, which explains the ``plateau'' in the
behavior of ${\rm F}^{(0)}_{2d}(Q)$ shown in Fig.~\ref{fF2d}(b). It is
important, however, that a similar calculation of the contribution to $F_{2d}$
coming from the photovoltaic term $d_s^{(D)}$ exactly cancels the constant
term in Eq.~(\ref{asymptote}), so that the difference between the two falls
off rapidly with increasing $Q$ for $Q\gg\alpha^{-1}$, as observed in
Fig.~\ref{fF2d}(a).

To summarize, in the limit of strongly overlapping Landau levels and low
temperatures, $1\ll\alpha\ll\tau_{\rm in}/\tau_{\rm q}$, there appear five
regions of $Q$ with essentially different behavior of the photoconductivity as
a function of $Q$ and $T$:
\begin{itemize} 
\item{$Q\ll\tau_{\rm q}/\tau_{\rm in}$: the photoresponse is linear in $Q$
and is strongly dependent on $T$;} 
\item{$\tau_{\rm q}/\tau_{\rm in}\ll Q\ll 1/\alpha$: the
strong feedback effect leads to the saturation of the photoresponse as a
function of $Q$. In this interval of $Q$, as well as at higher $Q$, the 
photoconductivity ceases to depend on $T$; }
\item{$\alpha^{-1}\ll Q\ll 1$: the feedback effects strongly suppress the
photoresponse;} 
\item{$1\ll Q\ll \alpha^2$: the multiphoton effects become important
and modify the feedback effects;} 
\item{$\alpha^2\ll Q$: the photoresponse is dominated by the multiphoton
effects, the feedback effects can be treated perturbatively. In the limit
$Q\to\infty$ the classical Drude conductivity is restored.}  
\end{itemize}

Above, in the bulk of Sec.~\ref{ss73}, we focused on the case of circular
polarization in the limit of high power. Two main differences of linear
polarization as compared to circular polarization are as follows. First, the
photoresponse becomes anisotropic, i.e., depends on the mutual orientation of
${\bE}_\omega$ and ${\bE}_{\rm dc}$. Second, the photoresponse may show much
stronger resonant features at $n\omega=2m\omega_c$, where $m$ and $n$ are
integer numbers. The latter is related to the different structure of
perturbation in $(\nu,n)$ space [Eq.~(\ref{df})] induced by the radiation. For
the case of passive circular polarization considered above, only the harmonics
$F_{2m,2m}$ along the diagonal are excited in the absence of the dc field
[they are all contained in the function $g_0$ in Eq.~(\ref{g0})]. The linear
response to the dc field, included in ${\rm St}_\bot$, couples these harmonics
with their neighbors $F_{2m,2m\pm 1}$. Under the action of ${\rm St}_j$ all
these harmonics contribute to $F_{\pm 1,0}$ and thus to the dc
photoconductivity. In the case of linear polarization, however, all harmonics
$F_{2m, n}$, with $m$ and $n$ integer, are excited by radiation, which leads
to the ``spreading'' of the perturbation all over the $(\nu,n)$ plane. The
resonances at $n\omega=2m\omega_c$ for the case of linear polarization warrant
additional study.

\section{Microwave--induced magnetoresistivity oscillations} 
\label{s8}
\setcounter{equation}{0}

In Secs.~\ref{s5}--\ref{s7} we have calculated the photoconductivity tensor
$\hat{\sigma}_{\rm ph}$ and discussed its evolution with increasing
``effective'' microwave power $Q$ defined in Eq.~(\ref{Q}) as
\be
Q=P\sin^2\frac{\pi\omega}{\wc}\,\omega^2
\left[\,{s_+^2\over
(\omega_c+\omega)^2}+{s_-^2\over (\omega_c-\omega)^2}\,\right]~,
\label{QQ}
\ee
where 
\be
P=\frac{2\tau_{\rm q}}{\tau_{\rm tr}}
\left(\frac{e E_\omega v_F}{\omega^2}\right)^2~.
\label{P}
\ee
However, what is usually measured in the experiment is the photoresistivity
$\hat{\rho}_{\rm ph}$ as a function of $\omega_c$ at fixed $\omega$ and $P$.
The dependence of $\hat{\rho}_{\rm ph}$ on $\omega_c/\omega$ for a given $P$
is discussed below, assuming, as in Secs.~\ref{ss72} and \ref{ss73}, passive
circular polarization of the microwave field.

For circular polarization, the photoresistivity tensor is easily obtained in
the limit of classically strong magnetic field, $\omega_c\tau_{\rm tr}\gg 1$,
by inverting Eq.~(\ref{cirph}):
\bea
\hat{\rho}_{\rm ph}&\simeq&
{1\over 2\sigma^{\rm D}_{xx}(\omega_c\tau_{\rm tr})^2}
\nonumber\\
&\times&
\left(\!\begin{array}{cc}1+2\delta^2(1-d_s)&\wc\tau_{\rm tr}-2\delta^2 h_a\\ 
-\wc\tau_{\rm tr}+2\delta^2
h_a&1+2\delta^2(1-d_s)\end{array}\!\right)~.\nonumber\\
\label{rhoph}
\eea
Here we neglect a small admixture of $-4\delta^2 h_a/\wc\tau_{\rm tr}$ in the
diagonal part of $\hat{\rho}_{\rm ph}$, as well as a similar admixture of $\pm
4\delta^2 d_s/\wc\tau_{\rm tr}$ in the Hall part.

\begin{figure}[ht]
\centerline{ 
\includegraphics[width=\columnwidth]{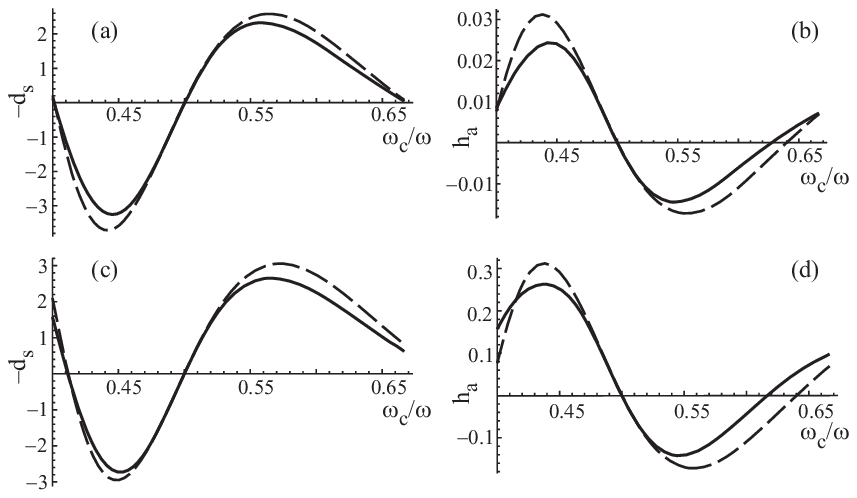}
}
\caption{Diagonal, $d_s$, and Hall, $h_a$, parts of the
  magnetoresistivity tensor (\ref{rhoph}) at moderate microwave power
  $P=0.01$ [Eq.~(\ref{P})] vs $\wc/\omega$. Solid lines: the
  functions $-d_s$ [panels (a) and (c)] and $h_a$ [panels (b) and
  (d)]; (a) and (b) correspond to $\alpha_{1/2}=1$, while (c) and (d)
  to $\alpha_{1/2}=10$.  Dashed lines: the linear--in--$P$ asymptotes
  $d_s^{(B)}+d_s^{(D)}$ and $h_a^{(D)}$ [Eqs.~(\ref{dsB1}),
  (\ref{dsD1}), and (\ref{ha1})]. }
\label{0.01}
\end{figure}

\begin{figure}[ht]
\centerline{ 
\includegraphics[width=\columnwidth]{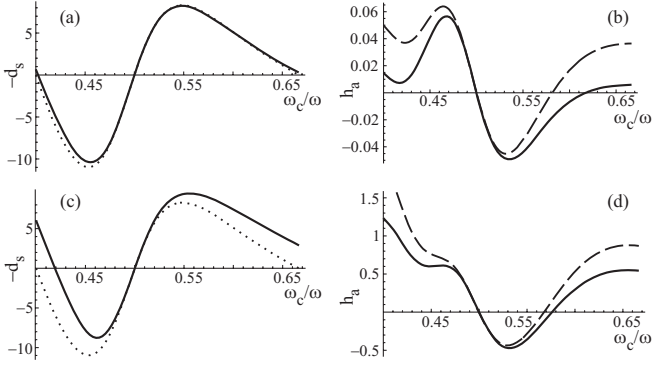}
}
\caption{Solid lines: same as in Fig.~\ref{0.01} for $P=0.1$.
Dashed lines: the asymptote for $h_a$ in the SIC regime [Eq.~(\ref{haha})].
Dotted lines: the saturated inelastic contribution [Eq.~(\ref{dsB})].  }
\label{0.1}
\end{figure}

\begin{figure}[ht]
\centerline{ 
\includegraphics[width=\columnwidth]{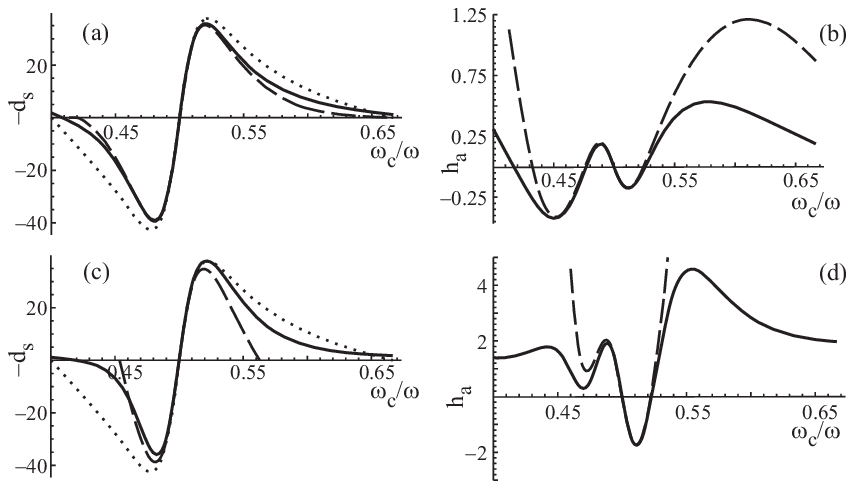}
}
\caption{Solid lines: same as in Fig.~\ref{0.1} for $P=1$.  Dashed
  lines: the asymptotes for $d_s$ and $h_a$ in the SIC regime
  [Eqs.~(\ref{dss}) and (\ref{haha})].  Dotted lines: the saturated
  inelastic contribution [Eq.~(\ref{dsB})].}
\label{1}
\end{figure}

\begin{figure}[ht]
\centerline{ 
\includegraphics[width=\columnwidth]{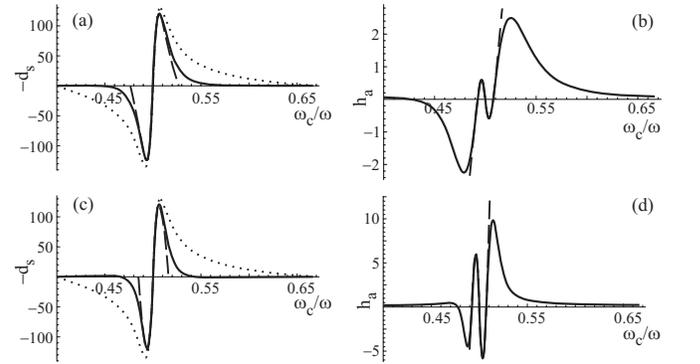}
}
\caption{Same as in Fig.~\ref{1} for $P=10$.}
\label{10}
\end{figure}

It was shown in Sec.~\ref{s7} that, while in the linear regime with respect to
the microwave power the inelastic contribution to the photoresponse dominates,
already at $Q\sim\alpha^{-1}\sim 1$ the inelastic and photovoltaic
contributions become comparable in magnitude. The displacement contribution
becomes also relevant at this power if $\alpha\sim 1$. The interplay of the
three contributions in the traces of $d_s$ and $h_a$ as a function of the
ratio $\omega_c/\omega$ is illustrated in Figs.~\ref{0.01}--\ref{10} for the
case of passive circular polarization. The dependence of $d_s$ and $h_a$ on
$\wc$ is shown over a single period of oscillations around the second harmonic
of the cyclotron resonance for $2/5<\wc/\omega<2/3$. Each of
Figs.~\ref{0.01}--\ref{10} corresponds to one of four values of the microwave
power $P=0.01,\,0.1,\,1$, and 10.

Within each of Figs.~\ref{0.01}--\ref{10} two cases are illustrated,
corresponding to two different values of the parameter $\alpha$
[Eq.~(\ref{alpha})] taken at $\wc/\omega=1/2$. In every figure, (a) and (b)
correspond to $\alpha_{1/2}=(3\wc\tau_{\rm q})^{-1}=1$, while (c) and (d) to
$\alpha_{1/2}=10$. In all the figures we take the same ratio $\tau_{\rm
in}/\tau_{\rm q}=50$, thus neglecting the heating effect, discussed separately
in Sec.~\ref{s92}.

At $P=0.01$ (Fig.~\ref{0.01}), not only $Q$ but also $Q_{\rm
  in}=\tau_{\rm in} Q/2\tau_{\rm q}$ is small compared to unity in the
whole interval $2/5<\wc/\omega<2/3$. Therefore, the photoresponse is
well described by the linear-in-$P$ asymptotes
(\ref{ds1})--(\ref{ha1}). More specifically, the diagonal part at
$\alpha\sim 1$ is dominated by the inelastic contribution $d_s^{(B)}$
[Eq.~(\ref{dsB1})], while at $\alpha\gg 1$ the photovoltaic
contribution $d_s^{(D)}$ [Eq.~(\ref{dsD1})] becomes pronounced near
the ends of the interval (where inelastic contribution vanishes, $\cot
w =0$). The Hall response is governed by the photovoltaic contribution
$h_a^{(D)}$ [Eq.~(\ref{ha1})]. The displacement part $d_s^{(A)}$ is
small due to the large ratio $\tau_{\rm in}/\tau_{\rm q}=50$, while
the quadrupole contribution to the Hall part, $h_s^{(C)}$
[Eq.~(\ref{hs1})], as well as the anisotropic diagonal part
(\ref{da1}) are absent for circular polarization. In the case of
linear polarization, $h_s^{(C)}$ would provide a Hall contribution of
the same order of magnitude as $h_a^{(D)}$. For comparison, the linear
asymptotes $d_s^{(B)}+d_s^{(D)}$ and $h_a^{(D)}$ are shown in
Fig.~\ref{0.01} by the dashed lines. The mismatch between the solid
and dashed lines is negligible for small detuning from the resonance,
\be
\label{Delta}
\Delta\equiv|\omega/\wc -2|\ll 1~,
\ee
owing to the $\sin^2\pi\Delta$ term in Eq.~(\ref{QQ}). Away from the
resonance, at $\Delta\sim1/2$, the parameter $Q_{\rm in}\simeq 0.1$
and the feedback effects (Sec.~\ref{s6}) lead to a noticeable
correction to the linear behavior.

The evolution of $d_s$ and $h_a$ with increasing $P$ reveals the nonlinear
effects studied in Secs.~\ref{s6} and \ref{s7}. First of all, note that the
part of $\hat{\sigma}_{\rm ph}$ associated with ${\rm F}_{1d}$ and ${\rm
F}_{1h}$ [Eqs.~(\ref{dsF}) and (\ref{haF})] vanishes both at $\Delta=0$ and at
$\wc/\omega=2/5$ and $2/3$, where $\Delta=1/2$. The photoresponse at both ends
of the interval is, therefore, fully governed by the contributions to
$\hat{\sigma}_{\rm ph}$ coming from ${\rm F}_{2d}$, ${\rm F}_{3d}$, ${\rm
F}_{2h}$ and ${\rm F}_{3h}$, which represent subleading terms for small
$P$. In accord with the results of Sec.~\ref{s7}, Figs.~\ref{0.1}(a) and (c)
show that the photoresponse at $\Delta=1/2$ in $d_s$ is indeed discernible
only at $P\agt\alpha^{-1}$; otherwise, it is masked by the much stronger
inelastic contribution that develops at $\Delta<1/2$.


In the vicinity of the resonance, the
diagonal photoresistivity is dominated by the $F_{1d}$ term in
Eq.~(\ref{dsF}), which at $\Delta^2\ll (P\alpha)^{-1}$ can be
represented as
\bea
\nonumber
d_s\simeq w \cot w \,{\rm F}_{1d}
&\simeq& {2\pi\omega\over\wc}\frac{P^*_{\rm in}\sin(2\pi\omega/\wc)}
{1+P^*_{\rm in}\sin^2(\pi\omega/\wc)}\\
&-&{9\pi\omega\over2\wc}\,P^*\sin{2\pi\omega\over\wc}
\nonumber
\\
&-&{\pi\omega\over4\wc}\sin{2\pi\omega\over\wc}\left(\frac{P^*}
{\omega\tau_{\rm q}}\sin{\pi\omega\over\wc}\right)^2\nonumber\\
&\times&
\frac{(\omega+\wc)^2+3 \wc^2}{(\omega+2\wc)^2}~,\label{dss}
\eea
where
\be
\label{P*}
P^*=P\frac{\w^2}{(\omega+\wc)^2}~,\qquad
P^*_{\rm in}=\frac{\tau_{\rm in}}{2\tau_{\rm q}}P^*~,
\ee
see Eqs.~(\ref{dsBfull}), (\ref{dsABsat}), and (\ref{F1d}). The asymptote
(\ref{dss}) is shown in panels (a) and (c) of Figs.~\ref{1} and \ref{10} by
the dashed line, while the dotted line describes the behavior of the first
term of Eq.~(\ref{dss}), which represents the saturated inelastic contribution
[Eq.~(\ref{dsB}), see also Eq.~(15) of Ref.~\onlinecite{long}]. For $P\gg
\alpha^{-1}$ the photoresponse tends to concentrate at $\Delta\ll 1$, which
can be seen from its evolution with increasing $P$ in
Figs.~\ref{0.01}--\ref{10}.

Similarly, the Hall part at $\Delta^2\ll (\alpha P)^{-1}$ can be approximated
by
\bea
\nonumber
h_a&\simeq& \frac{2\wc P^*\sin^2(\pi\omega/\wc)}
{\omega\tau_{\rm q}(\omega+2\wc)}\\\nonumber
&+&2\pi\left(1+{\wc\over\omega}\right)
\left(\frac{P^*\sin^2(\pi\omega/\wc)}
{(\omega+2\wc)\tau_{\rm q}}\right)^2\\
&-&\frac{\pi P^*\sin(2\pi\omega/\wc)}
{(\omega+2\wc)\tau_{\rm q}}\;
\frac{1-P^*_{\rm in}\sin^2(\pi\omega/\wc)}
{1+P^*_{\rm in}\sin^2(\pi\omega/\wc)}~,\nonumber\\
\label{haha}
\eea
see Eqs.~(\ref{haDB}), (\ref{F2h}), and (\ref{QAI}). The asymptote
(\ref{haha}) is shown in panels (b) and (d) of Figs.~\ref{0.1}--\ref{10} by
the dashed line.

Let us now briefly recall the physics behind the calculated dependencies by
the example of Fig.~\ref{10}, where the microwave power $P=10$ is large and
the effective power $Q$ [Eq.~(\ref{QQ})] changes from $Q=0$ at
$\wc/\omega=1/2$ to $Q\sim 10$ at $\wc/\omega=2/5$ and $2/3$. As a result, in
Fig.~\ref{10}, the system passes with increasing $\Delta$ through all regimes
discussed in Secs.~\ref{s5}--\ref{s7}.

\begin{itemize}

\item{$\Delta=0$: exactly on resonance, the photoresponse is zero.}

\item{$\Delta\ll\Delta_{\rm max}\ll 1$: the linear photoresponse,
$d_s,\,h_a\propto P \Delta$ (Sec.~\ref{s5}). The inelastic contribution
dominates the photoresponse in the diagonal part,
\be
\label{dslinear}
d_s=-{8\over9}\,(2\pi)^2\,{\tau_{\rm
  in}\over\tau_{\rm q}}\,P\Delta
\ee
[linear--in--$P$ part of the first term in Eq.~(\ref{dss})], while in the Hall
part the photovoltaic mechanism produces
\be
\label{halinear}
h_a=-{1\over 3}(2\pi)^2\,\alpha_{1/2} \,P\Delta~,
\ee
[linear--in--$P$ part of the last term in Eq.~(\ref{haha})]. The maximum value
of the inelastic term in $d_s$ as a function of $\Delta$ is reached at
\be
\label{DeltaMax} 
\Delta_{\rm max}\equiv\sqrt{9\tau_{\rm q}/2\pi^2\tau_{\rm in} P}~.
\ee
}
\item{$\Delta\sim\Delta_{\rm max}$: the crossover to the SIC regime
(Sec.~\ref{s6}). As detuning from the resonance $\Delta$ approaches
$\Delta_{\rm max}$, the nonlinear corrections in the first term of
Eq.~(\ref{dss}) and in the last term of Eq.~(\ref{haha}), driven by the
feedback effects related to the strong oscillatory pattern in the isotropic
part $F_{00}(\varepsilon)$ of the distribution function, become
important. At $\Delta=\Delta_{\rm max}$, the inelastic contribution to $d_s$
reaches maximum, $d_s=\pm 4\Delta_{\rm max}^{-1}\propto \sqrt{P}$, and crosses
over into the $\Delta^{-1}$ decay at $\Delta>\Delta_{\rm max}$ (illustrated by
the dotted lines). In the Hall term $h_a$, the strong oscillations of $F_{00}$
change sign of the photovoltaic contribution at $\Delta=\Delta_{\rm max}$. 
}
\item{$\Delta_{\rm max}\ll\Delta\ll (P\alpha)^{-1/2}:$ the SIC regime, in
which the photoresponse becomes independent of $\tau_{\rm in}$ and, hence, on
temperature. This property of the high--power photoresponse can be easily
verified experimentally.

At $\Delta\gg\Delta_{\rm max}$, the two last terms in Eq.~(\ref{dss}) become
relevant, leading to a decrease of the diagonal photoresponse.  The last term
is actually irrelevant in Fig.~\ref{10}(a) (which corresponds to
$\alpha_{1/2}=1$), so that [Eq.~(\ref{F1d})]:
\bea
d_s\simeq\frac{\pi\omega}{\wc}\cot\frac{\pi\omega}{\wc}(4-9Q)
\simeq{8\over\Delta}-8\pi^2 P\Delta~.  
\label{dsimple}
\eea
The second term in Eq.~(\ref{dsimple}) consists of the displacement
contribution modified by the strong oscillations in $F_{00}$ [in the SIC
regime, its sign is inverted compared to the linear case, Eq.~(\ref{dsAB})]
and the two--photon correction to $F_{00}$ itself [Eq.~(\ref{phiB1})]. This
term is responsible for the strong narrowing of the oscillation in $d_s$ as a
function of $\Delta$ and also for the nonmonotonic dependence on power at
fixed $\Delta$. At $\alpha\gg 1$, the last, photovoltaic term in
Eq.~(\ref{dss}) becomes important, leading to additional narrowing of the peak
in Fig.~\ref{10}(c) and Fig.~\ref{1}(c) compared to Fig.~\ref{10}(a) and
Fig.~\ref{1}(a).

The Hall part $h_s$ exhibits an additional oscillation at $\Delta\sim
(P\alpha)^{-1/2}$ and decays for larger $\Delta$ due to the feedback
and multiphoton effects. At $\Delta\gg\Delta_{\rm max}$, the first two
terms in Eq.~(\ref{haha}) (which, in contrast to the last term, are
even in detuning $\omega-2\wc$ from the resonance) become pronounced,
as can be seen, for example, from the difference in amplitude of
the second peak and the second dip in Fig.~\ref{10}(d).}

\item{$\Delta\gg (P\alpha)^{-1/2}$: the ultrahigh power regime, in which all
effects related to the Landau quantization decay to zero due to the feedback
and multiphoton effects.}

\end{itemize}

Figures~\ref{0.01}--\ref{10} demonstrate that, for large $P$ and/or large
$\alpha$, the strongly nonlinear effects studied in Sec.~\ref{s7} essentially
modify the behavior characteristic of the SIC regime\cite{long} (dotted lines
in Fig.~\ref{0.1}--\ref{10}). In particular, high--power measurements should
reveal the significant narrowing of the oscillation in $\rho_{xx}$ with
increasing $P$, manifest in Figs.~\ref{1} and \ref{10}. It should be noted
that, in addition to the magnetoresistivity traces dominated by the strong
feature in the vicinity of $\Delta=0$, it is worth studying the $Q$ and
$T$--dependences of the photoresponse at several fixed ratios $\omega/\wc$; in
particular, the contributions ${\rm F}_{2d}$ and ${\rm F}_{3d}$ which manifest
themselves at $\Delta\sim 1/2$.

At $P\sim\alpha^{-1}$ the magnitude of the photoresponse in $\rho_{xy}$,
Fig.~\ref{0.1}(d), becomes comparable to the effect in $\rho_{xx}$,
Fig.~\ref{0.1}(c), which seems to be in accord with the
experiment.\cite{maniHall} However, the experiments that revealed the Hall
oscillations with $h_a\simeq -d_s$ were performed at $\omega\tau_{\rm q}\sim
1$, i.e., at $\alpha\sim 1$, where for the level of power used in the
experiment the predicted oscillations in $h_a$ should be more than an order of
magnitude smaller than those in $d_s$. Our theory is not capable of explaining
the observed effect in $\rho_{xy}$ (probably related to microwave--induced
$\omega/\wc$--dependent changes in the electron concentration, see discussion
in Sec.~VIII of Ref.~\onlinecite{long}).

\section{Range of applicability of the theory}
\label{s9}
\setcounter{equation}{0}

The above calculation the OPC is based on the assumptions that (i) disorder is
smooth, Eq.~(\ref{disorder}); (ii) temperature is high, Eq.~(\ref{fosc});
(iii) Landau levels strongly overlap with each other,
Eq.~(\ref{nuOvLL}). Altogether, these conditions made the analytical treatment
of the problem possible to all orders in the microwave power. Now let us
briefly discuss the range of applicability of the above approximations,
especially in the limit of strongly nonlinear photoresponse. In particular, it
is important to discuss the role of heating. 

\subsection{Heating effects}
\label{s91}

The high--$T$ approximation led us to Eq.~(\ref{fosc}) for the
oscillatory distribution function (see Appendix~\ref{A1}). The
Fermi--Dirac part of the distribution is characterized by the
effective temperature $T_e\gg\omega_c$. The oscillatory part appears
due to the Landau quantization, which, in the limit of strongly
overlapping Landau levels, is weak: the amplitude of the oscillations
of the DOS is proportional to the small parameter $\delta$
[Eq.~(\ref{nuOvLL})]. The heating of electrons, which leads to a
growth of $T_e$ with increasing microwave power, can therefore be
treated separately within the framework of the classical Boltzmann
equation with a constant DOS. Such a calculation leads to the
following equation for $T_e$:\cite{classical}
\bea
\label{heating}
&&(T_e/T_0)^4\,(T_e/T_0-1)=\eta~,\\
&&\eta={\tau_{\rm e-ph}(T_0)\over\tau_{\rm tr}} 
\,\left(\frac{e v_F E_\omega}{2 T_0}\right)^2
\,\sum\limits_\pm\left(\frac{s_\pm}{\omega\pm\wc}\right)^2~.\nonumber\\
\label{eta}
\eea
Here $\tau_{\rm e-ph}^{-1}(T_0)\propto T_0^3$ is the inelastic relaxation rate
due to scattering on acoustic phonons (the latter represent a thermal bath at
temperature $T_0$) and $s_\pm$ describes polarization of microwaves
[Eq.~(\ref{E})]. At a typical value of $T_0\sim 1$ K, the electron--phonon
scattering rate exceeds $\tau_{\rm tr}^{-1}$ by more than an order of
magnitude, so that the parameter $\eta$ in Eq.~(\ref{eta}) can vary between 1
and 1000 under realistic experimental conditions. At first glance, the large
values of $\eta$ imply a strong heating, $T_e\gg T_0$. However, this is not
actually the case in view of the strong temperature dependence of $\tau_{\rm
e-ph}(T)$. While at $\eta\lesssim 1$ the temperature grows linearly with
$\eta$ (specifically, $T_e-T_0\simeq\eta$), in the limit of strong heating the
growth is much weaker, $T_e/T_0\simeq\eta^{1/5}\gg 1$. The weak dependence on
the microwave power at large $\eta$ means that, in practice, the heating can
be substantial but never strong, i.e., $T_e/T_0\sim 2\div3$. At the same time,
with lowering bath temperature well below $1\,$K, while keeping the microwave
power constant, the heating effect may become significantly more pronounced
(since $\eta$ grows with lowering temperature as $T_0^{-5}$), in effect
restoring the high--$T$ limit we deal with in the paper.

In the dc response, the strongest manifestation of the heating effect is the
exponential suppression of the Shubnikov-de Haas oscillations (provided the
latter are visible in the absence of microwaves at $T_e=T_0$). The OPC is
influenced much more weakly, since $T_e$ enters the photoconductivity only
indirectly, through the $T$--dependence of the electron-electron inelastic
scattering rate, $\tau_{\rm in}^{-1}\propto T_e^2$. In particular, in the
linear regime, $P\ll \tau_{\rm q}/\tau_{\rm in}$, taking the heating effects
into account leads to a sublinear dependence of the diagonal photoresponse on
the microwave power, namely $d_s\propto PT_e^{-2}(P)$, while at high power,
$P\gg \tau_{\rm q}/\tau_{\rm in}$, the maxima and minima of $d_s$ scale as
$\sqrt{P}/T_e(P)$.

\subsection{Stratification of the distribution function}
\label{s92}

Apart from the condition on the temperature $T_e \gg 1/t_B$, the approximation
(\ref{fosc}) is valid only if the oscillatory correction to the distribution
function remains much smaller than unity, so that the general requirement
$0<f(\ve)<1$ is met. Despite the DOS being only weakly modulated ($\delta\ll
1$), the above condition is violated at sufficiently high power for the values
of the ratio $\omega/\omega_c$ corresponding to the minima and maxima of $d_s$
[see Figs.~\ref{10}(a) and (c)]. Near these points, the effect is governed by
the inelastic mechanism, Eq.~(\ref{dsimple}), and the oscillatory part of
Eq.~(\ref{fosc}) can be estimated as
\be
\label{stra} 
{\cal F}(\ve)\,
\partial_\ve f_T\sim \frac{ \delta \,d_s}{t_B\, T_e}\sin(\ve t_B)
\alt\frac{ \delta }{t_B\, T_e \,\Delta_{\rm  max}}~,
\ee
where we used Eqs.~(\ref{prephi}), (\ref{defphibot}), (\ref{phiB}),
(\ref{dsB}), and (\ref{dsimple}). 

\begin{figure}[ht]
\centerline{ 
\includegraphics[width=0.95\columnwidth]{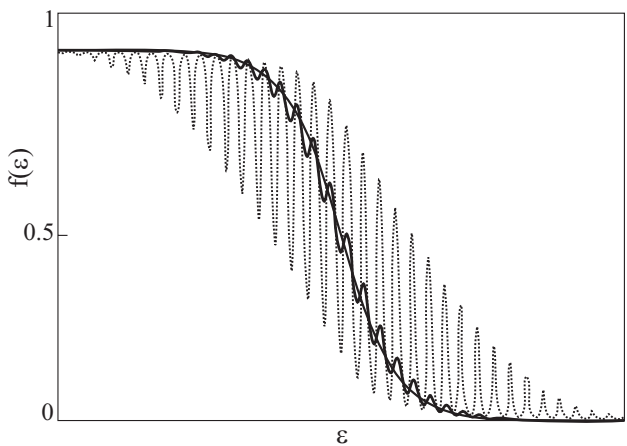}
}
\caption{Isotropic part of the distribution function at three levels
  of the microwave power: $P=0$, $P\ll P_\star$, and $P\sim P_\star$.
}
\label{stratification}
\end{figure}

At large power, the maximum value of $d_s$ grows as $\sqrt{P}/T_e(P)$
[see Eq.~(\ref{dslinear}) at $\Delta=\Delta_{\rm max}$]. It follows
that at $P\sim P_\star=(\wc\delta/T_e)^2\tau_{\rm in}/\tau_{\rm q}$
the oscillations of the isotropic part of the distribution near the
maxima and minima of $d_s$ become of order unity. At small detuning
from the cyclotron resonance harmonics, $\Delta\sim\Delta_{\rm max}$,
and at $P>P_\star$, the splitting of $f(\ve)$ into the smooth and
oscillatory parts (\ref{fosc}) is no longer possible. The distribution
function in this regime can be calculated numerically according to
Eq.~(11) of Ref.~\onlinecite{long} (for details of the numerical
procedure, see Ref.~\onlinecite{crossover}). An illustrative example
of the stratification of the distribution function at $P\sim P_\star$
is shown in Fig.~\ref{stratification}. Both the heating and the
stratification of the distribution function suppress the growth and
the narrowing of the peak in $d_s$ at very high power. Beyond the
close vicinity of maxima and minima, at
$|\Delta|=|\omega/\wc-N|\gg\delta/t_B\, T_e$ (where $N$ is integer),
the oscillatory correction to the distribution function remains small
at any power level and the high--$T$ approximation works well, so that
our results remain intact whatever the microwave power.

\subsection{Smooth vs short--range disorder}
\label{s93}

In this paper we assumed that the disorder potential is created solely by
charged impurities separated from the plane of the 2DEG by a spacer of width
$\xi/2\gg k_F^{-1}$ [Eq.~(\ref{disorder})]. Such a minimal model of disorder
allowed us to account for the experimentally relevant small--angle scattering
condition $\tau_{\rm q}\gg\tau_{\rm tr}$ and to consider in an unambiguous way
the region of magnetic fields $\tau_{\rm tr}^{-1}\ll\wc\ll \tau_{\rm
q}^{-1}$. On the other hand, a ``two--component model'' of
disorder,\cite{twocomponent} including strong scatterers on the background of
the smooth potential (\ref{disorder}), is known to provide a better
description of real ultrahigh--mobility structures.
 
To our current understanding, the inclusion of the strong scatterers into the
theory should not lead to any qualitative change of the results presented here
[provided the conditions (\ref{parameters}) remain satisfied]. In particular,
the derivation of the linear-order inelastic contribution (\ref{dsB1}) in
Ref.~\onlinecite{dmitriev03} does not require any assumptions as to the type
of disorder.

By contrast, the strong scatterers were
shown\cite{yang02,2componentDC} to play a crucial role in the
nonlinear dc response (in the absence of microwaves), where strong
oscillations of the magnetoresistivity (``HIRO'') were
observed,\cite{yang02,bykov1,bykov2,inelasticDC,strongDC} governed by
the ratio of the Hall electric field to the magnetic field.
Remarkably, a strong interplay between two types of oscillations in
the 2DEG driven by both microwave and strong dc fields was recently
observed experimentally.\cite{ACDC} Theoretical description of this
experimental situation necessitates the inclusion of the
backscattering off strong impurities, which we relegate to future
work.  It was argued in Ref.~\onlinecite{Volkov} that, in the regime
of strong dc field and in the presence of short range disorder,
effects of the microwave field on screening of impurities should also
be taken into account.

\subsection{Stronger magnetic fields, $\wc\tau_{\rm q}\agt 1$}
\label{s94}

The calculations in Secs.~\ref{s5}-\ref{s7} were performed for the case of
strongly overlapping Landau levels, $\wc\tau_{\rm q}\ll 1$. In this limit, (i)
microwave-induced corrections to the DOS can be neglected; (ii) there are
substantial effects related to excitation of higher angular and temporal
harmonics of the distribution function; in particular, the photovoltaic and
quadrupole contributions to the OPC. At stronger magnetic fields, the
modulation of the DOS due to the Landau quantization becomes more pronounced:
in the SCBA approximation, at $\wc\tau_{\rm q}\simeq 1.8$ the Landau levels
get separated from each other.\cite{crossover} For the case of separated
Landau levels, the effect on higher harmonics of the distribution function
becomes strongly suppressed and the diagonal part of the photoresponse is
determined by the interplay of the inelastic and displacement mechanisms.

In addition to the nonlinear effects considered in this paper, at
$\wc\tau_{\rm q}\agt 1$ the DOS itself is modified by microwave
radiation. In particular, microwave--induced sidebands appear on both
sides of the Landau levels, leading to the appearance of an additional
structure in the OPC pattern near the fractional harmonics of the
cyclotron resonance, $n\omega=m\wc$.\cite{separated} A similar
structure in photoresponse also arises as a multiphoton correction to
the inelastic contribution to the OPC. Both structures have precisely
the same shape (as a function of $\omega$); however, the multiphoton
mechanism dominates in the case of strongly separated Landau
levels.\cite{separated} Such a ``fractional oscillatory pattern''
(FOP) in the nonlinear photoresponse was observed in recent
experiments\cite{zudov04,multi,dorozhkin06} and explained in terms of
either of multiphoton corrections to the displacement
contribution\cite{zudov04,multi,leiliumulti} or, alternatively, in
terms of single--photon resonant corrections to the inelastic
contribution which are specific to the crossover region $\wc\tau_{\rm
q}\sim 1$.\cite{dorozhkin06,crossover} All four contributions to the
FOP will be compared and analyzed elsewhere.\cite{separated}

\section{Conclusions}
\label{s10}
\setcounter{equation}{0}

Summarizing, we have developed the systematic approach to the
microwave--induced oscillations in the magnetoresistivity of a 2DEG. This
approach has enabled us to classify contributions to the photoresistivity
according to the combined action of the microwave and dc fields on the
temporal and angular harmonics of the distribution function
(Sec.~\ref{s3}). We have studied the interplay of the resulting mechanisms of
photoresponse at high microwave power. In the limit of strongly overlapping
Landau levels (Sec.~\ref{s4}), the dc photoconductivity has been calculated to
all orders in the microwave power (Sec.~\ref{s5}--\ref{s7}).

To linear order in the power, two novel mechanisms of the oscillations
(quadrupole and photovoltaic) have been identified, in addition to the known
ones (displacement\cite{ryzhii} and inelastic\cite{dmitriev03}). The
quadrupole and photovoltaic mechanisms have been shown to be the only ones
leading to oscillations in the Hall part of the photoconductivity tensor. Of
particular interest is the result that the quadrupole contribution violates
Onsager symmetry. In the diagonal part, the inelastic contribution dominates
at moderate microwave power, while at elevated power the contributions of
other mechanisms become important.

In Secs.\ref{s6} and \ref{s7} we have considered the strongly nonlinear
photoresponse at high microwave power. We have shown that a competition
between various nonlinear effects (the feedback effects, the excitation of
high angular and temporal harmonics of the distribution function, and the
multiphoton effects) drives the system through four different nonlinear
regimes with increasing microwave power.

Most dramatic changes in the photoresponse are due to the feedback effects. In
the SIC regime, Sec.\ref{s6}, the feedback from the microwave--induced
oscillations of the isotropic part of the distribution, $F_{00}$, leads to
the saturation of the inelastic contribution,\cite{short,long} and to the strong
interplay of the inelastic effect and all other contributions to the OPC. In
particular, the strong oscillations of $F_{00}$ change sign of the
most relevant parts of the displacement and photovoltaic contributions.
Remarkably, in the SIC regime the photoresponse becomes independent
of the inelastic scattering rate and, therefore, of temperature.

At higher power, Sec.~\ref{s7}, the feedback suppresses the effects on higher
temporal and angular harmonics of the distribution function. At still higher
power, the multiphoton excitation becomes pronounced and starts to compete
with the feedback effects. At ultrahigh power, the feedback and multiphoton
effects destroy all quantum contributions, restoring the classical Drude
conductivity.

Our theory predicts nonmonotonic behavior of the diagonal photoresponse as a
function of the microwave power at a fixed ratio $\omega/\wc$. As illustrated
in Sec.~\ref{s8}, in magnetoresistivity measurements such a nonmonotonic
dependence should result in a significant narrowing of the oscillatory
structure around the integer values of $\omega/\wc$ at sufficiently large
power, see Eq.~(\ref{dsimple}). To a large extent, the narrowing is due to the
displacement contribution. At the same time, the amplitude of the oscillations
is controlled by the inelastic mechanism at any power level. We suggest to
experimentally measure the power and temperature dependences of the
photoresponse at fixed $\omega/\omega_c$ not too close to the cyclotron
resonance harmonics. Such experiments will make it possible to observe the
rich behavior of the photovoltaic contributions. Also, our theory predicts
oscillations in $\rho_{xy}$, which, in the limit of strongly overlapping
Landau levels, can be comparable in amplitude with those in $\rho_{xx}$. 

We thank S.I.Dorozhkin, R.R.~Du, K.~von~Klitzing, J.H.~Smet, and
especially M.A.~Zudov for information about the experiments,
D.N.~Aristov, A.P.~Dmitriev, and I.V.~Gornyi for stimulating
discussions, and E.E.~Takhtamirov and V.A.~Volkov for sending us the
paper Ref.~\onlinecite{Volkov} prior to publication. This work was
supported by the SPP ``Quanten-Hall-Systeme'' and Center for
Functional Nanostructures of the DFG, by INTAS Grant
No.~05-1000008-8044, and by the RFBR.

\appendix
\section{Quantum Boltzmann equation in the high-temperature limit}
\label{A1}
\renewcommand{\theequation}{A.\arabic{equation}}
\setcounter{equation}{0}

In this Appendix we derive the high--$T$ version of the quantum Boltzmann
equation starting from its general form, Eqs.~(\ref{kineq}), (\ref{St_im}). We
show that the solution of the QBE in the limit $T\gg\wc$ has the form
\be\label{f21}
f_{21}\equiv f(t_-,\,t)={\cal G}(t_-)+\sum\limits_{n=-\infty}^{\infty}
\Phi_n(\varphi,t)\,it_-^{(n)}{\cal G}(t_-^{(n)})
\ee
and obtain the equations for the amplitudes $\Phi_n(\varphi, t)$.
Here $t_-=t_2-t_1$, $t=(t_2+t_1)/2$, and 
${\cal
G}(t_-)=(2\pi)^{-1}\int\! d\ve\, f_T(\ve)\exp(-i\ve t_-)$ is the Fermi distribution function in the time representation,
\[
{\cal G}(t_-)=\frac{i\,T\exp(-i \ve_F t_-)}{2 \sinh \pi T (t_-+i0)}
\,,\qquad t_-^{(n)}=t_--n t_B~.
\]

In the energy representation, the solution (\ref{f21}) acquires the form
(\ref{fosc}),
\bea
f(\ve,\varphi,t)&=&\int\! d t_- \, e^{i\ve t_-}f_{21}\nonumber\\
&=&f_T(\ve)+{\cal F}(\ve,t,\varphi)\,\partial_\ve f_T(\ve)~,\\
\label{F}
{\cal F}(\ve,\varphi,t)&=&
\sum\limits_{n=-\infty}^{\infty} \Phi_n(\varphi,t) \,\exp(i n \ve  t_B)~.
\eea  
In the high--$T$ limit, the function $f(\ve)$ shows fast oscillations on top
of the smooth thermal distribution $f_T(\ve)$.  The function $f_{21}$,
Eq.~(\ref{f21}), is composed of equidistant sharp peaks of width $1/T$,
with the distance between them being equal to $t_B$. 

The DOS (\ref{nu}) for high Landau levels is a periodic function of energy and
the functions $g^R=-\left(g^A\right)^\dagger$, entering Eq.~(\ref{St_im}),
obey
\be
\label{g21}
g^R_{21}=\sum\limits_{m=0}^\infty\,g_m(t)\,\delta(t_-^{(n)})~.
\ee 
In terms of the coefficients $g_m$, the QBE is rewritten as 
\be
\label{kineqA}
(\partial_t+\wc\partial_\varphi)f_{31}
-\St_{\rm in}\{f\}_{31}=\St_{\rm im}\{f\}_{31}
\ee
with
\begin{widetext}
\bea
\nonumber
\St_{\rm im}\{f\}_{31}&=&\sum\limits_{m\geq 0} \left[\, \hat{\cal K}_{31}g_m
\left(\!\frac{t_2+t_3}{2}\!\right)f_{21}
- f_{21}\hat{\cal K}_{32}g_m
\left(\!\frac{t_2+t_3}{2}\!\right)\,\right]_{t_2=t_3-m t_B} \\
\label{StimA}
&+&
\sum\limits_{m\geq 0}\left[\, \hat{\cal K}_{31}f_{32}g_m^*
\left(\!\frac{t_2+t_1}{2}\!\right)
-f_{32}\hat{\cal K}_{21}g_m^*
\left(\!\frac{t_2+t_1}{2}\!\right)\,\right]_{t_2=t_1-m t_B}~. 
\eea 
Now we substitute the solution in the form (\ref{f21}) into the kinetic
equation.  The unperturbed part ${\cal G}(t_-)$ gives zero on the l.h.s.\ of
Eq.~(\ref{kineqA}). The impurity collision integral acting on ${\cal G}(t_-)$,
produces
\bea
\nonumber 
 \St_{\rm im}\{{\cal G}\}_{31}
&=&\sum\limits_{m\geq 0}\left.  {\cal G}(t_2-t_1)[ \hat{\cal K}_{31}-\hat{\cal
K}_{32}] g_m(t_3-m t_B/2)\right|_{t_3-t_2=m t_B}\\\label{StG}
&+&\sum\limits_{m\geq 0}\left. {\cal G}(t_3-t_2)[ \hat{\cal K}_{31}-\hat{\cal
K}_{21}] g_m^*(t_1-m t_B/2)\right|_{t_1-t_2=m t_B}~,
\eea 
where we took into account the fact that the operator $\hat{\cal K}$, in the
time representation, commutes with an arbitrary function independent of
$\varphi$, see Eq.~(\ref{K}).
\end{widetext}
Let us consider the first sum in the above expression. At $T t_B\gg
1$, the $\delta$-function-like ${\cal G}(t_2-t_1)$ puts $t_2$ within
the interval $|t_2-t_1|\alt 1/T$. Strictly at the point $t_2-t_1=0$,
the function ${\cal G}(t_2-t_1)$ is infinite, while the expression in
the square brackets is zero. Recalling that the kernel changes
smoothly on the scale of $|t_2-t_1|\sim 1/T\ll t_B$, one can replace
the difference in the brackets by the derivative $\partial_1\hat{\cal
K}_{31}$. The result is a series of peaks of identical shape,
\bea
\nonumber 
&&\St_{\rm im}\{{\cal G}\}_{21} =\sum\limits_{-\infty}^\infty{\cal A}_N(t)
\,it_-^{(n)}{\cal G}(t_-^{(n)})~,\\
\nonumber 
&&{\cal A}_{N> 0}=-i\left[\,(\partial_{t_-}-\partial_{t}/2) \hat{\cal
K}_{21}\,\right]_{t_-=N t_B}\,g_N(t)~,\\
\nonumber 
&&{\cal A}_{N< 0}=-i\left[\,(\partial_{t_-}+\partial_{t}/2) \hat{\cal
K}_{21}\,\right]_{t_-=N t_B}\,g^*_{-N}(t)~,\\
&&{\cal A}_{N=0}= -2i\left[\,\partial_{t_-}\hat{\cal K}_{21}\,
\right]_{t_-=0}g_0=
-i\left.\partial_{t_-}\hat{\cal K}\right|_{t_-=0}~,\nonumber\\
\label{A}
\eea 
where the square brackets mean that the time derivatives act on the kernel
only and do not act on the functions $g_N(t)$. In the last expression, we took
into account that $g_0$, representing the DOS in the absence of magnetic
field, is not affected by the external fields and by disorder: the average DOS
is a conserved quantity, $g_0=1/2$.

One can see that $\St_{\rm im}\{{\cal G}\}$, given by Eq.~(\ref{A}), has
exactly the same form as the oscillatory correction to the distribution
function, Eq.~(\ref{f21}). It follows that the solution at $T t_B\gg1$ indeed
has the form (\ref{f21}). Substituting the perturbed part of the distribution
function (\ref{f21}) into the kinetic equation, Eqs.~(\ref{kineqA}), and
(\ref{StimA}), we finally arrive at the equations for the amplitudes $\Phi_n$,
\bea
\nonumber
&&(\partial_t+\wc\partial_\varphi)\Phi_N+\tau_{\rm in}^{-1}
\langle \Phi_N\rangle
={\cal A}_N\\\nonumber
&&+\sum\limits_{m\geq 0}\left[
\hat{\cal K}_{N}(t)\Phi_{N-m}(t_m)g_m(t_{m-N})\right.\\\nonumber
&&-\Phi_{N-m}(t_m)\hat{\cal K}_{m}(t_{m-N})g_m(t_{m-N})\\\nonumber
&&+\hat{\cal K}_{N}(t)\Phi_{N+m}(t_m)g^*_m(t_{m+N})\\
\label{highT}
&&\left.
-\Phi_{N+m}(t_m)\hat{\cal K}_{-m}(t_{m+N})g^*_m(t_{m+N})\right]~,
\eea
where $\hat{\cal K}_{n}(t_m)=\hat{\cal K}(t_-=n t_B,\,t=t_m)$ and $t_m=t-m
t_B/2$. The inelastic relaxation, which controls the magnitude of the
oscillations in the isotropic part of the distribution function, can be
described in the relaxation time approximation, with $\tau_{\rm in}$ being the
effective electron--electron inelastic scattering time.\cite{long}

\section{Quantum Boltzmann equation: Overlapping Landau levels}
\label{A2}
\renewcommand{\theequation}{B.\arabic{equation}}
\setcounter{equation}{0}

In this appendix we consider the high--$T$ version of the QBE,
Eq.~(\ref{highT}), in the limit of weak magnetic field, which corresponds to
the exponentially small modulation of the DOS. In this case, only first two
terms in the expansion (\ref{g21}) should be taken into account, namely,
$g_0=1/2$, and $g_1=-\delta$, where $\delta=\exp(- t_B/2\tau_q)\ll 1$. Other
terms, $g_n={\cal O}(\delta^n)$, are exponentially smaller and can be
neglected. We will also use the smallness of the odd part of the kernel,
$\hat{\cal K}_j$, with respect to $\hat{\cal K}_\bot$, controlled by the
parameter $\sqrt{\tau_{\rm q}/\tau_{\rm tr}}$. At zero order in $\hat{\cal
K}_j$, the distribution function is even in $\phi$ and the current is
zero. Thus, $\hat{\cal K}_j\ll\hat{\cal K}_\bot$ should be taken into account
once, while the contributions of the second and higher orders in $\hat{\cal
K}_j$ should be neglected.

At leading order in $\delta$, the kinetic equation (\ref{highT}) gives
\be
\label{F00}
(\partial_t+\wc\partial_\varphi)\Phi_0=
-i\left.\partial_{t_-}\hat{\cal K}\right|_{t_-=0}=
-i\left.\partial_{t_-}\hat{\cal K}_j\right|_{t_-=0}~,
\ee
where the operator $\hat{\cal K}$ at the last position in any expression
should be understood as acting on unity. The inelastic term on the l.h.s.\ is
omitted as the resulting $\Phi_0$ contains odd angular harmonics only. Indeed,
according to Eq.~(\ref{X}), $\X(t_-=0)=0$. It follows that only odd part of
the kernel, $\hat{\cal K}_j$, produces a non-zero contribution to the r.h.s.\
of Eq.~(\ref{F00}) [see Eqs.~(\ref{K})--(\ref{Kfi})]. Also, on the r.h.s.\ of
Eq.~(\ref{F00}) we omitted
\be
\label{term}
[\hat{\cal K}_{0}(t)\Phi_{0}(t)-\Phi_{0}(t)\hat{\cal K}_0(t)]/2=
\tau_{\rm tr}^{-1}\partial_\varphi^2 \Phi_0~,
\ee
originating from the $m=0$ term in Eq.~(\ref{highT}). (Here we used again the
fact that $\X(t_-=0)=0$, which gives $\hat{\cal K}_{0}=\tau_{\rm
q}^{-1}+\left.\hat{\cal K}_\varphi\right|_{\X=0}$). In the case of classically
strong magnetic field, $\wc\tau_{\rm tr}\gg1$, the term (\ref{term}) can be
safely neglected.  Note, however, that the dc conductivity is infinite in the
absence of this term in the opposite limit, $B=0$.

At order ${\cal O}(\delta)$, the kinetic equation (\ref{highT}) produces an
oscillatory correction with the amplitude $\Phi_1(\varphi,t)$ obeying
\bea
\nonumber
&&(\partial_t+\wc\partial_\varphi)\Phi_1(t)+\tau_{\rm in}^{-1}\langle \Phi_1\rangle
=
i\delta(\partial_{ t_B}-\partial_{t}/2)
\hat{\cal K}(t_B,t)\\
&&+[\,\hat{\cal K}_1(t)-\hat{\cal K}_0\,]\,\Phi_1(t)-\delta\,[\,\hat{\cal
K}_1(t)\Phi_0(t_1)-\Phi_0(t_1)\hat{\cal K}_1(t)\,]~.
\nonumber\\
\label{F1}
\eea
In classically strong magnetic field, the last term should be neglected, since
(i) $\Phi_0$, according to Eq.~(\ref{F00}), is generated by $\hat{\cal
K}_j\propto \sqrt{\tau_{\rm q}/\tau_{\rm tr}}\ll 1$; (ii) the leading part of
the kernel, $\hat{\cal K}_\bot\propto (\tau_{\rm q}/\tau_{\rm tr})^0$,
commutes with $\varphi$. Thus, only the odd part of the kernel, $\hat{\cal
K}_j$, produces a nonzero term in the curly brackets, and the result, of
second order in $\hat{\cal K}_j$, is proportional to $\tau_{\rm q}/\tau_{\rm
tr}$ and should be neglected. With the same accuracy, ${\cal O}(\sqrt{\tau_{\rm
q}/\tau_{\rm tr}})$, the expression in the square brackets should be replaced
by $[\,\hat{\cal K}_j(t)+\hat{\cal K}_\bot(t)\,]_{t_-= t_B}$.

In the high-$T$ limit, the leading contribution to the current is of order
$\delta^2$ [Sec.~\ref{s4}]. Accordingly, we represent the nonoscillatory part
of the solution (\ref{F}) as
\be
\label{F0}
\Phi_0=\phi_0^{(D)}+2\delta^2\phi_0^{(2)}~,
\ee
with $\phi_0^{(D)}$ obeying Eq.~(\ref{F00}) and $\phi_0^{(2)}$ generated by
$m=1$ term in the sum (\ref{highT}),
\bea
\nonumber
&&2\delta(\partial_t+\wc\partial_\varphi)\phi_0^{(2)}(t)\\
&&=
-\hat{\cal K}_0(t)\Phi_{-1}(t_1)+\Phi_{-1}(t_1)\hat{\cal K}_1(t_1\nonumber)\\
&&-\hat{\cal K}_0(t)\Phi_{1}(t_1)+\Phi_{1}(t_1)
\hat{\cal K}_{-1}(t_1)~.
\label{FF02}
\eea
Now we recall that the distribution function (\ref{F}) is real,
$\Phi_n=\Phi_{-n}^*$, $\hat{\cal K}_n=\hat{\cal K}^*_{-n}$, and take into account
that ${\rm Im}\hat{\cal K}=-i\hat{\cal K}_j$. After that, Eq.~(\ref{FF02}) can
be rewritten as
\bea
\nonumber
&&(\partial_t+\wc\partial_\varphi)\phi_0^{(2)}(t)=\\
&&{\rm Im}\,[\phi_1(t_1)]i\hat{\cal K}_j(t_B, t_1)
-{\rm Re}\,[\phi_1(t_1)]\hat{\cal K}_\bot(t_B, t_1)~,\nonumber\\
\label{F02}
\eea
where the part $\hat{\cal K}_\varphi$ of the kernel is neglected, $t_1=t_B/2$,
and we switched to the more convenient notation of Sec.~\ref{s4} by putting
$\phi_1=-\Phi_1/\delta$.

Equation (\ref{F1}), being rewritten separately for the real and imaginary
parts, reads
\bea
&&(\partial_t+\wc\partial_\varphi){\rm Im}\phi_1(t)+\tau_{\rm in}^{-1}
\langle{\rm Im} 
\phi_1(t)\rangle
\nonumber\\
&&=(\partial_{t}/2-\partial_{ t_B})\hat{\cal K}_\bot(t_B,t)
+\hat{\cal K}_\bot(t_B,t){\rm Im}\phi_1(t)~,
\nonumber\\
\label{ImF1}\\
&&(\partial_t+\wc\partial_\varphi){\rm Re}\phi_1(t)=
(\partial_{t}/2-\partial_{ t_B})i\hat{\cal K}_j(t_B,t)
\nonumber\\
&&+\hat{\cal K}_\bot(t_B,t){\rm Re}\phi_1(t)
+i\hat{\cal K}_j(t_B,t){\rm Im}\phi_1(t)~.
\label{ReF1}
\eea
Equation (\ref{ReF1}) shows that ${\rm Re}\phi_1$ is of first order in
$\hat{\cal K}_j$, hence an odd function of $\varphi$. Accordingly, we omitted
the inelastic collision term in Eq.(\ref{ReF1}), as well as the term
$i\hat{\cal K}_j(t_B, t_1){\rm Re}\phi_1(t)$ on the r.h.s.\ of Eq.(\ref{ImF1})
which is of second order in $\hat{\cal K}_j$ and thus small in parameter
$\tau_{\rm q}/\tau_{\rm tr}$.

The similarity of the r.h.s.\ of Eq.~(\ref{ReF1}) to that of Eq.~(\ref{F02})
is not accidental. In fact, the dc current is expressed in terms of the
combined quantity, $\phi_j(t)=\phi_0^{(2)}(t+t_B/2)+{\rm Re}\phi_1(t)$, see
Eq.~(\ref{Ovcur}). Both Eqs.~(\ref{ReF1}) and (\ref{F02}) contain the term
$\hat{\cal K}_\bot(t_B, t){\rm Re}\phi_1(t)={\rm Re}\phi_1(t)\hat{\cal
K}_\bot(t_B, t)$ [unlike $\hat{\cal K}_j$, the operator $\hat{\cal K}_\bot$
commutes with $\varphi$, see Eqs.~(\ref{Kbot}), and (\ref{Kj})]. This term
generates contributions to $\phi_0^{(2)}$ and ${\rm Re}\phi_1(t)$, as
illustrated by diagrams (E) and (F) in Fig.~\ref{diagrams}. However, in the dc
current (\ref{Ovcur}) these contributions cancel each other, as the r.h.s.\ of
the equation for $\phi_j$ contains $\hat{\cal K}_j$ only, whereas
$\phi_\bot(t)={\rm Im}\phi_1(t)$ is fully governed by $\hat{\cal K}_\bot$.
The final equations for $\phi_0^{(D)}$, $\phi_\bot$, and $\phi_j$,
Eqs.~(\ref{kinD})--(\ref{kinbot}), are given in Sec.~\ref{s4}.


\begin{thebibliography}{1}

\bibitem[*]{byline} Also at A.F.~Ioffe Physico-Technical Institute,
194021 St.~Petersburg, Russia.

\bibitem[$\dagger$]{} Also at Petersburg Nuclear Physics
Institute, 188350 St.~Petersburg, Russia.

\bibitem{zudov01} M.A.~Zudov, R.R.~Du, J.A.~Simmons, and J.R.~Reno,
cond-mat/9711149; Phys.\ Rev.\ B {\bf 64}, 201311(R) (2001);

\bibitem{Ye01} P.D.~Ye, L.W.~Engel, D.C.~Tsui, J.A.~Simmons, J.R.~Wendt,
G.A.~Vawter, and J.L.~Reno, Appl.\ Phys.\ Lett.\ {\bf 79}, 2193 (2001).

\bibitem{yang02} C.L.~Yang, J.~Zhang, R.R.~Du, J.A.~Simmons, and J.L.~Reno,
Phys.\ Rev.\ Lett.\ {\bf 89}, 076801 (2002).

\bibitem{mani02} R.G.~Mani, J.H.~Smet, K.~von~Klitzing, V.~Narayanamurti,
W.B.~Johnson, and V.~Umansky, Nature {\bf 420}, 646 (2002).

\bibitem{zudov03} M.A.~Zudov, R.R.~Du, L.N.~Pfeiffer, and K.W.~West, Phys.\
Rev.\ Lett.\ {\bf 90}, 046807 (2003).

\bibitem{yang03} C.L.~Yang, M.A.~Zudov, T.A.~Knuuttila, R.R.~Du,
L.N.~Pfeiffer, and K.W.~West, Phys.\ Rev.\ Lett.\ {\bf 91}, 096803 (2003).

\bibitem{dorozhkin03} S.I.~Dorozhkin, JETP Lett.\ {\bf 77}, 577 (2003).

\bibitem{willett03} R.L.~Willett, L.N.~Pfeiffer, and K.W.~West,
Phys.\ Rev.\ Lett.\ {\bf 93}, 026804 (2004).

\bibitem{B-periodic} I.V.~Kukushkin, M.Yu.~Akimov, J.H.~Smet, S.A.~Mikhailov,
K.~von~Klitzing, I.L.~Aleiner, V.I.~Falko, Phys.\ Rev.\ Lett.\ {\bf 92},
236803 (2004).

\bibitem{zudov04} M.A.~Zudov, Phys.\ Rev.\ B {\bf 69}, 041304(R) (2004).

\bibitem{du04}R.R.~Du, M.A.~Zudov, C.L.~Yang, Z.Q.~Yuan, L.N.~Pfeiffer, and
K.W.~West, Int.\ J. Mod.\ Phys.\ B {\bf 18}, 3465 (2004).

\bibitem{mani04} R.G.~Mani, J.H.~Smet, K.~von~Klitzing,
  V.~Narayanamurti, W.B.~Johnson, and V.~Umansky, Phys.\ Rev.\ B {\bf
    69}, 193304 (2004); Phys.\ Rev.\ Lett.\ {\bf 92}, 146801 (2004);
  R.G.~Mani, V.~Narayanamurti, K.~von~Klitzing, J.H.~Smet,
  W.B.~Johnson, and V.~Umansky, Phys.\ Rev.\ B {\bf 70}, 155310
  (2004).

\bibitem{maniHall}R.G.~Mani, V.~Narayanamurti, K.~von~Klitzing, J.H.~Smet,
W.B.~Johnson, and V.~Umansky, Phys.\ Rev.\ B {\bf 69}, 161306(R) (2004).

\bibitem{Studenikin04} S.A.~Studenikin, M.~Potemski, P.T.~Coleridge,
A.~Sachrajda, and Z.R.~Wasilewski, Solid State Commun.\ {\bf 129}, 341 (2004).

\bibitem{Kovalev} A.E.~Kovalev, S.A.~Zvyagin, C.R.~Bowers, J.L.~Reno, and
J.A.~Simmons, Solid State Commun.\ {\bf 130}, 379 (2004).

\bibitem{mani05} R.G.~Mani, Appl.\ Phys.\ Lett.\ {\bf 85}, 4962 (2004);
Phys.\ Rev.\ B {\bf 72}, 075327 (2005).

\bibitem{polarization} J.H.~Smet, B.~Gorshunov, C.~Jiang, L.~Pfeiffer,
  K.~West, V.~Umansky, M.~Dressel, R.~Meisels, F.~Kuchar, and
  K.~von~Klitzing, Phys.\ Rev.\ Lett.\ {\bf 95}, 116804 (2005).

\bibitem{studenikin05} S.A.~Studenikin, M.~Potemski, A.~Sachrajda, M.~Hilke,
  L.N.~Pfeiffer, and K.W.~West, Phys.\ Rev.\ B {\bf 71}, 245313 (2005).

\bibitem{studenikin06} S.A.~Studenikin, M.~Byszewski, D.K.~Maude,
  M.~Potemski, A.~Sachrajda, Z.R.~Wasilewski, M.~Hilke, L.N.~Pfeiffer,
  and K.W.~West, Physica E {\bf 34}, 73 (2006); S.A.~Studenikin,
  M.~Byszewski, D.K.~Maude, M.~Potemski, A.~Sachrajda, M.~Hilke,
  L.N.~Pfeiffer, and K.W.~West, cond-mat/0602079.

\bibitem{bykov1} A.A.~Bykov, J.-Q.~Zhang, S.~Vitkalov, A.K.~Kalagin, and
  A.K.~Bakarov, Phys.\ Rev.\ B {\bf 72}, 245307 (2005).

\bibitem{bykov2} A.A.~Bykov, A.K.~Kalagin, A.K.~Bakarov, and A.I.~Toropov,
  JETP Lett.\ {\bf 81}, 348 (2005); {\it ibid.} {\bf 81}, 406 (2005).

\bibitem{parr B} C.L.~Yang, R.R.~Du, L.N.~Pfeiffer, and K.W.~West, Phys.\
Rev.\ B {\bf 74}, 045315 (2006).

\bibitem{antidot} Z.Q.~Yuan, C.L.~Yang, R.R.~Du, L.N.~Pfeiffer, and K.W.~West,
Phys.\ Rev.\ B {\bf 74}, 075313 (2006).

\bibitem{dorozhkinINTRA} S.I.~Dorozhkin, J.H.~Smet, V.~Umansky, and
K.~von~Klitzing, Phys.\ Rev.\ B {\bf 71}, 201306(R) (2005).


\bibitem{bichrom} M.A.~Zudov, R.R.~Du, L.N.~Pfeiffer, and K.W.~West, Phys.\
Rev.\ Lett.\ {\bf 96}, 236804 (2006).

\bibitem{multi} M.A.~Zudov, R.R.~Du, L.N.~Pfeiffer, and K.W.~West, Phys.\
Rev.\ B {\bf 73}, 041303(R) (2006).


\bibitem{dorozhkin06} S.I.~Dorozhkin, J.H.~Smet, K.~von~Klitzing,
L.N.~Pfeiffer, and K.W.~West, cond-mat/0608633.


\bibitem{inelasticDC} J.-Q.~Zhang, S.~Vitkalov, A.A.~Bykov, A.K.~Kalagin,
and A.K.~Bakarov, Phys.\  Rev.\  B {\bf 75}, 081305(R) (2007).


\bibitem{strongDC} W.~Zhang, H.-S.~Chiang, M.A.~Zudov, L.N.~Pfeiffer, and
K.W.~West, Phys.\ Rev.\ B {\bf 75}, 041304(R) (2007).

\bibitem{ACDC} W.~Zhang, M.A.~Zudov, L.N.~Pfeiffer, and K.W.~West,
  Phys.\ Rev.\ Lett. {\bf 98}, 106804 (2007).


\bibitem{andreev03} A.V.~Andreev, I.L.~Aleiner, and A.J.~Millis, Phys.\ Rev.\
Lett.\ {\bf 91}, 056803 (2003).

\bibitem{ryzhii} V.I.~Ryzhii, Sov.\ Phys.\ Solid State {\bf 11}, 2078 (1970);
  V.I.~Ryzhii, R.A.~Suris, and B.S.~Shchamkhalova, Sov.\ Phys.\ Semicond.\
  {\bf 20}, 1299 (1986).

\bibitem{durst03} A.C.~Durst, S.~Sachdev, N.~Read, and S.M.~Girvin, Phys.\
Rev.\ Lett.\ {\bf 91}, 086803 (2003).

\bibitem{dmitriev03} I.A.~Dmitriev, A.D.~Mirlin, and D.G.~Polyakov,
  Phys.\ Rev.\ Lett.\ {\bf 91}, 226802 (2003).

\bibitem{Raikh03} A.A.~Koulakov and M.E.~Raikh, Phys.\ Rev.\ B {\bf 68},
  115324 (2003).

\bibitem{shi} J.~Shi and X.C.~Xie, Phys.\ Rev.\ Lett.\ {\bf 91}, 086801
(2003).

\bibitem{ryzhii03} V.~Ryzhii and R.~Suris, J. Phys.: Condens.\ Matter {\bf
    15}, 6855 (2003); V.~Ryzhii, Phys.\ Rev.\ B {\bf 68}, 193402 (2003);
    V.~Ryzhii and V.Vyurkov, {\it ibid.} {\bf 68}, 165406 (2003).

\bibitem{leiliu03} X.L.~Lei and S.Y.~Liu, Phys.\ Rev.\ Lett.\ {\bf 91}, 226805
  (2003).

\bibitem{VA} M.G.~Vavilov and I.L.~Aleiner, Phys.\ Rev.\ B {\bf 69}, 035303
  (2004).

\bibitem{short} I.A.~Dmitriev, M.G.~Vavilov, I.L.~Aleiner, A.D.~Mirlin, and
D.G.~Polyakov, Physica E {\bf 25}, 205 (2004).

\bibitem{classical} I.A.~Dmitriev, A.D.~Mirlin, and D.G.~Polyakov,
Phys.\ Rev.\ B {\bf 70}, 165305 (2004).

\bibitem{compress} M.G.~Vavilov, I.A.~Dmitriev, I.L.~Aleiner, A.D.~Mirlin, and
D.G.~Polyakov, Phys.\ Rev.\ B {\bf 70}, 161306(R) (2004).

\bibitem{michailov04} S.A.~Mikhailov, Phys.\ Rev.\ B {\bf 70}, 165311 (2004).

\bibitem{park} K.~Park, Phys.\ Rev.\ B {\bf 69}, 201301(R) (2004).

\bibitem{lee} D.-H.~Lee and J.M.~Leinaas, Phys.\ Rev.\ B {\bf 69}, 115336
(2004).

\bibitem{Raikh04} C.~Joas, M.E.~Raikh, and F.~von~Oppen, Phys.\ Rev.\ B {\bf
70}, 235302 (2004).

\bibitem{Lyapilin1} I.I.~Lyapilin and A.E.~Patrakov, cond-mat/0510260.

\bibitem{ryzhii04} V.~Ryzhii, A.~Chaplik, and R.~Suris, JETP Lett.\ {\bf 80},
363 (2004).

\bibitem{falkoprl} J.P.~Robinson, M.P.~Kennett, N.R.~Cooper, and V.I.~Fal'ko,
Phys.\ Rev.\ Lett.\ {\bf 93}, 036804 (2004).

\bibitem{long} I.A.~Dmitriev, M.G.~Vavilov, I.L.~Aleiner, A.D.~Mirlin, and
D.G.~Polyakov, Phys.\ Rev.\ B {\bf 71}, 115316 (2005).

\bibitem{Halperin} A.~Auerbach, I.~Finkler, B.I.~Halperin, and A.~Yacoby,
Phys.\ Rev.\ Lett.\ {\bf 94}, 196801 (2005).

\bibitem{Balents} J.~Alicea, L.~Balents, M.P.A.~Fisher, A.~Paramekanti, and
L.~Radzihovsky, Phys.\ Rev.\ B {\bf 71}, 235322 (2005).

\bibitem{vonoppen1} J.~Dietel, L.I.~Glazman, F.W.J.~Hekking, and F.~von~Oppen,
Phys.\ Rev.\ B {\bf 71}, 045329 (2005).

\bibitem{vonoppen2} C.~Joas, J.~Dietel, and F.~von~Oppen, Phys.\ Rev.\ B {\bf
72}, 165323 (2005).

\bibitem{torres05} M.~Torres and A.~Kunold, Phys.\ Rev.\ B {\bf 71}, 115313
  (2005).

\bibitem{leiliu05} X.L.~Lei and S.Y.~Liu, Appl.\ Phys.\ Lett.\ {\bf 86},
  262101 (2005); Phys.\ Rev.\ B {\bf 72}, 075345 (2005).

\bibitem{Volkov} E.E.~Takhtamirov and V.A.~Volkov, cond-mat/0506727;
  JETP {\bf 104}, 602 (2007).

\bibitem{michailov04a} S.A.~Mikhailov and N.A.~Savostianova, Phys.\ Rev.\ B
{\bf 71}, 035320 (2005); {\it ibid.} {\bf 74}, 045325 (2006).

\bibitem{falkoprb} M.P.~Kennett, J.P.~Robinson, N.R.~Cooper, and V.I.~Fal'ko,
Phys.\ Rev.\ B {\bf 71}, 195420 (2005).

\bibitem{dietel} J.~Dietel, Phys.\ Rev.\ B {\bf 73}, 125350 (2006).

\bibitem{torres06} M.~Torres and A.~Kunold, J. Phys.: Condens.\ Matter {\bf
18}, 4029 (2006).

\bibitem{Kashuba06} A.~Kashuba, Phys.\ Rev.\ B {\bf 73}, 125340 (2006); JETP
Lett.\ {\bf 83}, 293 (2006).

\bibitem{Lyapilin2} I.I.~Lyapilin and A.E.~Patrakov, cond-mat/0606311.

\bibitem{leiliumulti} X.L.~Lei and S.Y.~Liu, Appl.\ Phys.\ Lett.\ {\bf 88},
212109 (2006).

\bibitem{leiliubichrom} X.L.~Lei and S.Y.~Liu, Appl.\ Phys.\ Lett.\ {\bf 89},
182117 (2006).

\bibitem{2componentDC} M.G.~Vavilov, I.L.~Aleiner, and L.I.~Glazman,
cond-mat/0611130.

\bibitem{Auerbach} A.~Auerbach and G.V.~Pai, cond-mat/0612469.

\bibitem{crossover} I.V.~Pechenezhskii, S.I.~Dorozhkin, and I.A.~Dmitriev,
  JETP Lett.\ {\bf 85}, 86 (2007). 


\bibitem{Fitzgerald} R.~Fitzgerald, Physics Today {\bf 56}, 24 (2003).

\bibitem{Girvin} A.C.~Durst and S.M.~Girvin, Science {\bf 304}, 1752 (2004).

\bibitem{Lyap} I.I.~Lyapilin and A.E.~Patrakov, Low Temp.\ Phys.\ {\bf 30},
  834 (2004). 

\bibitem{Durst06} A.C.~Durst, Nature {\bf 442}, 752 (2006).


\bibitem{tauin} The inelastic time $\tau_{\rm in}$ was experimentally
  determined in Ref.~\onlinecite{inelasticDC} and was found to be in agreement with the
  theoretical prediction of Ref.~\onlinecite{long}. It should be noted, however, that
  the theoretical conditions $\omega_c\ll T$, $\omega_c\tau_q\ll 1$
  were only marginally satisfied in the experiment [\onlinecite{inelasticDC}].


\bibitem{Ando} T.~Ando and Y.~Uemura, J. Phys.\ Soc.\ Jpn.\ {\bf 36}, 959
  (1976); T.~Ando, A.B.~Fowler, and F.~Stern, Rev.\ Mod.\ Phys.\ {\bf 54}, 437
  (1982).

\bibitem{RaSha} M.E.~Raikh and T.V.~Shahbazyan, Phys.\ Rev.\ B {\bf 47}, 1522
  (1993).

\bibitem{twocomponent} A.D.~Mirlin, D.G.~Polyakov, F.~Evers, and P.~W\"olfle,
Phys.\ Rev.\ Lett.\ {\bf 87}, 126805 (2001).

\bibitem{antionsager} E.M.~Epshtein, Prog. Quantum Electron. {\bf 22}, 259 (1979).

\bibitem{separated} I.A.~Dmitriev, A.D.~Mirlin, and D.G.~Polyakov, to be
published.

\end{thebibliography}
\end{document}